# A systematic review of sample size determination in Bayesian randomized clinical trials: full Bayesian methods are rarely used

Yanara Marks[1], Jessie Cunningham[2], Arlene Jiang[1], Linke Li[1,3], Yi-Shu Lin[1], Abigail McGrory[1,3], Yongdong Ouyang[1], Nam-Anh Tran[4], Yuning Wang[1,3] and Anna Heath[1,3,5]

October 22, 2025

[1]Child Health Evaluative Sciences, Peter Gilgan Centre for Research and Learning, The Hospital for Sick Children

[2]Hospital Library and Archives, The Hospital for Sick Children

[3]Dalla Lana School of Public Health, University of Toronto

[4]Department of Epidemiology, Biostatistics and Occupational Health, School of Population and Global Health, Faculty of Medicine and Health Sciences, McGill University

[5]Department of Statistical Science, University College London

## Abstract

**Background:** Utilizing Bayesian methods in clinical trials has become increasingly popular, as they can incorporate prior information into the design, and allow for smaller sample sizes while providing reliable and robust statistical results. Various Bayesian methods for sample size determination are available, and while these methods are well justified and understood, it is unclear how they are being used in practice. This study aims to understand how sample sizes for Bayesian randomized clinical trials (RCTs) are determined and inform future designs of Bayesian trials.

**Methods:** A systematic literature review was conducted in May 2023 and updated in July 2025. We included completed RCTs which a) assessed the efficacy of interventions in humans; b) utilized a Bayesian framework for the primary data analysis; c) published in English; and d) enrolled participants between December 2009 – July 2025.

**Results:** The literature search produced 74,833 records, of which 27,890 were duplicates, and 46,943 were screened using manual and automated screening. 283 full texts were screened and 164 studies moved to extraction. Our findings demonstrate a slow increase in RCTs using Bayesian methods to analyse primary efficacy data from 2012 onwards, with a sharp increase during the COVID-19 pandemic (42%). The most common method for sample size determination in Bayesian RCTs was a hybrid approach (60%) in which elements of Bayesian and frequentist theory are combined. Bayesian RCTs predominantly took place in North America (34%) and mainly focused on adult study populations (85%). Bayesian trials were used in a variety of disease areas; the most common being COVID-19 (31%).

**Conclusion:** Fully Bayesian methods for sample size determination are rarely used in practice, despite significant theoretical development. Our review revealed a lack of standardized reporting across Bayesian RCTs, making it challenging to review the sample size determination. The CONSORT statement for reporting RCTs states that sample size calculations must be reported, which was poorly adhered to. Among RCTs that reported sample size determination, relevant information was frequently omitted from the reports and discussed in poorly structured supplementary materials. Thus, there is a critical need for greater transparency, standardization and translation of relevant methodology in Bayesian RCTs.

# 1 Introduction

Randomized controlled trials (RCTs) are widely regarded as the gold standard for the objective comparison of treatments, or interventions, within a given population, as they uphold methodological rigor and reduce the potential for bias.[1,2] Late-phase RCTs usually focus on evaluating whether a treatment improves an outcome of interest.[3] One crucial aspect of trial design is to determine a sample size that allows the trial to reach its objectives while providing valid ethical and scientific outcomes.[4–6] The exact method for sample size determination (SSD) will depend on the aims of the trial and the framework used for statistical inference.[7]

The frequentist framework is commonly used in efficacy RCTs  and makes inferences by considering the long-run frequency of events in an infinite number of identical clinical trials.[8] In this framework, SSD is performed by determining the number of participants required to have a sufficiently high chance of rejecting the null hypothesis if  an "important" difference between the interventions under investigation truly exists.[9–12] In this methodology, clinical input is required to determine a "minimally important clinical difference", which is the smallest difference between the interventions that would be considered clinically relevant.[13,14] To implement this approach, there is a large body of literature to support SSD for different outcomes and study designs.[15,16] These methods are implemented in standard software and online calculators and are well-established in clinical trials.[5]

More recently, Bayesian methodology in clinical trials has become increasingly popular. The Bayesian method provides an alternative definition of probability that relates to the degree of belief in a hypothesis.[17] Broadly, Bayesian inference combines trial data with information available before the trial, enumerated in a *prior distribution*, to quantify a posterior distribution that summarises all the available information about the treatment effect.[18] The Bayesian approach can offer several advantages.[19] Firstly, it can incorporate data from prior studies and expert opinions to potentially reduce the required sample size and costs.[4,20] Despite allowing for smaller sample sizes, the Bayesian approach provides equally reliable and robust statistical results.[4] This approach also aligns more closely to clinical decision making and can facilitate the use of complex adaptive designs.[19,21] Finally, Bayesian methods can be easier to implement in adaptive trials as they do not require adjustments for the interim analysis.[22]

Various methods have been proposed for Bayesian SSD, which reflect different perspectives on the goal of clinical trials.[5,23–32] Firstly, Bayesian SSD can use a hybrid approach where the analysis is performed from a Bayesian perspective, but the frequentist properties of the trial are calculated using fixed values for the quantities of interest.[18,33–35] A fully Bayesian extension to this method uses a prior distribution for these quantities, calculating a measure known as assurance or expected power.[36,37] Alternatively, fully Bayesian approaches can use a utility function or precision-based approaches.[5] While these approaches have been discussed in the statistical literature, there is limited research investigating practical SSD when utilizing Bayesian design and analysis in clinical trials.

To address this gap, we performed a systematic review of Bayesian clinical trials to determine how RCT sample sizes are calculated when using a Bayesian framework for the analysis of efficacy in clinical trials. We also aimed to understand how key inputs were selected for SSD.

## 2 Methods

### 2.1 Methods for Sample Size Determination

There are a range of methods for SSD, which take varied perspectives on the goal of an RCT. To present these methods, we will consider a 2-arm RCT with $n$ individuals randomized to each arm. The trial aims to determine whether a new intervention is more effective than the standard of care with the effect of treatment on the outcome of interest denoted $\theta$ by collecting data $\boldsymbol{X}$, which is summarized into an estimate for $\theta$, denoted $\hat{\theta}$.

### *2.1.1 Frequentist Methods*

In this setting, the frequentist approach would specify the null hypothesis $H_0: \theta = 0$ and the alternative hypothesis $H_1: \theta \neq 0$. This null hypothesis would be rejected if $|\hat{\theta}| \geq R$, where the exact value for $R$ depends on the distribution of $\hat{\theta}$, the variability of the individual-level outcome, and controls the probability of rejecting the null hypothesis, when it is true, below $\alpha$, where $\alpha$ usually equals 0.05.[38] The sample size of each arm in the proposed RCT is determined by assuming that $\theta = A$, where $A$ is a point value from the alternative hypothesis, and selecting a sample size such that

$$P\left(|\hat{\theta}| \geq R |\, \theta = A\right) > 1 - \beta,$$

where $\beta$ controls the probability that the null hypothesis is not rejected when it is false.[39] The choice of $A$ is critical and is usually defined as the smallest change in outcome that would be clinically meaningful.[40]

### *2.1.2 Hybrid Methods*

The frequentist SSD methodology is rooted in the frequentist definition of probability as it aims to control trial behavior over an infinite number of identical trials and assumes a fixed value for the parameter $\theta$. The hybrid approach to SSD for a Bayesian RCT aims to control the probability of drawing a definitive conclusion from the RCT, while redefining the method used to draw this conclusion.[41] Decision making in Bayesian RCTs are often based on posterior probabilities, i.e., computing $P(\theta > W|\boldsymbol{X})$, where $W$ could be 0 or an effect of interest and determining whether this probability is sufficiently high to conclude superiority of the novel intervention.[42] Trial conclusions can also use predictive probabilities, which calculate the probability of observing future data that falls within a certain rejection region,[42] or Bayes factors.[43]

### *2.1.3 Bayesian Methods*

Fully Bayesian SSD methods assume a distribution for all parameters and move away from the hypothesis testing framework. The simplest extension of frequentist methods assumes a "design" prior distribution for $A \sim p(A)$ that represents our current (informative) beliefs about the value of the treatment effect. The probability of declaring that the novel treatment is effective is then averaged over this prior distribution to calculate assurance,[24] or expected power.[25] The sample size is then estimated to provide a level of assurance, similar to the frequentist method.[26] Note that $p(A)$ does not need to match the prior distribution used in the primary analysis to compute the posterior distribution.[23]

An alternative perspective for Bayesian SSD aims to control the precision of the posterior distribution for $\theta$. Multiple methods have been proposed to achieve this including the average length criterion, the average coverage criterion or worst outcome criterion.[5,23] These methods aim to control the posterior credible interval, (the average interval or the worst possible interval), by either fixing the length, or coverage, and then increasing the sample size until a desired coverage, or length, is achieved. These methods were recommended for Bayesian SSD by Cao et al.[44] in 2009, which acts as the start date for our review. Recent reviews have also highlighted these methods.[27]

Finally, Bayesian SSD can use decision-theoretic approaches using utility theory. These approaches have been extensively developed in different contexts[29–32,45] but broadly aim to design trials by formally assigning a "utility" to each of the potential interventions and then assessing the impact of uncertainty on decision making. From this, we can determine the appropriate sample size by trading off the utility of deciding between the interventions based on additional data and based on the currently available data.[46]

#### *2.1.4 Adaptive Designs*

Bayesian methods are often used for adaptive trials where SSD is more complex and usually relies on simulation. This is because the sample size of the RCT changes depending on what data are collected during the trial. Nevertheless, for trial planning purposes calculations are often performed to evaluate the most likely sample sizes and the range of potential sample sizes to determine the trial time and budget. While reporting of these simulations can be complex, the expected sample size and the maximum sample size should still be available to confirm that the total recruitment is sufficient.

## 2.2 Systematic Review Methods

To evaluate which of these SSD methods are used, we conducted a systematic review following the Preferred Reporting Items for Systematic Reviews and Meta-Analysis (PRISMA)[47] guidelines. The study protocol was registered and can be accessed through the International Prospective Register of Systematic Reviews (PROSPERO) #CRD42023437915. We ran the initial search in May 2023, and conducted an updated search in January 2025, to include papers up to the end of 2024. This initial phase included manuscripts that discussed the sample size calculation in the title or abstract. We then expanded this initial search to include all published RCTs from December 2009 until July 2025.

### *2.2.1 Inclusion and Exclusion Criteria*

We included prospective RCTs that i) used Bayesian methodology for their primary data analysis, ii) had a primary data analysis focusing on the efficacy of interventions, iii) were completed, iv) were published in English, v) recruited human participants, and vi) started participant recruitment on/after December 2009.

We excluded: i) non-randomized or quasi-randomized clinical trials, ii) RCTs which relied on frequentist methodology for their primary data analysis, iii) papers that reported analyses other than primary analyses (i.e., post-hoc or secondary analyses), iv) incomplete or proposed trials that were not reporting a primary analysis, v) non-English publications, vi) conference abstracts or protocols vii) studies on animals, and viii) RCTs which had started participant recruitment prior to December 2009.

### *2.2.2 Detailed Research Methodology*

The search strategy was developed by a professional medical librarian (JC) and the search was executed in five bibliographic databases: Ovid MEDLINE(R), Ovid Embase, Embase Classic, EBM Revies, and SCOPUS, on May 16, 2023, with date limits of December 2009-2023, updated on January 16, 2025, and July 22, 2025. Results were uploaded to Covidence (Veritas Health Innovation, Melbourne, Australia), and duplicates automatically removed. The following Bayesian keywords were included to capture relevant publications: posterior probability, posterior distribution, prior distribution, prior, uninformative prior, credible interval, hierarchical model, dynamic borrowing, and predictive probability. See Supplementary Table 2 for the detailed search strategy.

### *2.2.3 Study Selection*

The first phase used a two-step selection process for abstract and full text screening to determine inclusion eligibility using Covidence systematic review software. Titles and abstracts were independently screened by two reviewers (NT, AH, YM, LL, AM, YL) and each study cataloged as "include", "exclude", or "maybe". Two reviewers collectively reviewed studies catalogued as "maybe" to reach consensus and resolve any discrepancies. Full texts were retrieved and screened independently by two reviewers (AJ, YM, LL, AM, YL, YW). Disagreements were discussed between the reviewers, and if not resolved, cross-checked by a third reviewer (AH) for final decision. The second phase was initially screened using a classifier trained on the data from the initial phase and built using the glmnet package in R. The classifier aimed to identify manuscripts that passed into full text screening and had an accuracy of 89% in the training set. Among the manuscripts that were included in the study, the classifier identified 94%. Titles and abstracts that passed this screening were then manually screened using the same process as above.

### *2.2.4 Data Extraction and Analysis*

Data were extracted by two teams, extracting independently using modified standardized extraction forms through Covidence. The first team (YM, YL) extracted study characteristics (author, publication year, country, trial phase, study design, objectives, clinical interventions and comparators, and primary outcomes), and population characteristics (disease area and distribution by age and sex). The second team (LL, AM, AJ, YO, YW) extracted methodological data, including maximum, expected, and actual sample size, and information on the SSD (framework, outcomes, prior distribution). Extracted data was checked for accuracy and completeness and verified by a reviewer who did not extract the data of the study (YO, AJ, AH). Elements that the team were unable to locate have been reported in the results and discussion sections.

### *2.2.5 Risk of Bias*

As this methodological review is analyzing Bayesian SSD for RCTs, a risk of bias analysis, assessing the quality of included studies, was not applicable. However, we examined whether trials reported their SSD as required by the CONSORT statement.[48,49]

## 3 Results

The initial literature search and update produced 19,182 records published between December 2009 and December 2024, from which 8,870 duplicates were removed, and 10,312 abstracts were screened. The extended search with date limits of December 1, 2009, until July 22, 2025, led to a total of 74,833 identified records, of which 27,890 duplicates were removed. The abstracts that were not previously screened (n=36,631) were screened using the automated screening. The automated screening identified an additional 1277 abstracts to be manually screened by our team. A total of 283 abstracts proceeded to full text screening, and 164 studies were selected for data extraction. Figure 1 displays the PRISMA flow diagram from the first, and updated literature searches.

### 3.1 Study Characteristics

Our results demonstrate a slow increase in the number of RCTs using a Bayesian approach to analyse their primary efficacy data between 2012 and 2019 (Figure 2). There was a sharp increase in Bayesian efficacy RCTs with 85% of trials published after 2020. This is likely due to the widespread use of Bayesian RCTs during the COVID-19 pandemic. The majority of RCTs were North American (55; 34%), followed by European (51; 31%) RCTs. The review also identified 35 (21%) international RCTs, followed by 6 (4%) RCTs carried out in Australia and South America, respectively. Lastly, 5 (3%) RCTs were conducted in Africa, 4 (2%) in Asia, and 2 (1%) in the Middle East. In this review, the transcontinental countries Türkiye and Russia were classified as part of Europe.

A variety of trial phases were reported, despite a focus on efficacy clinical trials. A total of 55 (34%) RCTs were phase 2, 52 (32%) were phase 3, 13 (8%) were seamless, combining two or more phases,[27] and 10 (6%) were phase 4. 34 (21%) RCTs did not specify a study phase, either because a traditional trial phase was not applicable, or the phase was not reported. Just less than half the RCTs (75; 46%) were adaptive. Finally, 109 (67%) RCTs randomized participants to two interventions, 31 (19%) to three interventions, and 22 (14%) to four or more interventions, with eight interventions being the maximum. Note that although one trial[50] used two arms, they randomized to four different treatment strategies within each arm. An additional two studies[51,52] reported on more than one randomization and were not included in this analysis.

### 3.2 Population Characteristics

Bayesian RCTs were used in many disease areas. The most common was COVID-19, with 51 (31%) RCTs, followed by oncology (16; 10%), neurology (14; 9%), cardiology (9; 5%), pulmonology (9; 5%), and infectious diseases (7; 4%). The remaining 58 (37%) RCTs looked at other disease areas, including perinatal health, psychology, rheumatology, and substance use (Table 1). Amongst these, we identified 20 (12%) RCTs looked at rare diseases.

Bayesian RCTs predominantly enrolled adults, with 139 (85%) RCTs enrolling participants aged 18 years or older, while 16 (10%) targeted mixed age groups, 8 (5%) recruited pediatric populations, defined as 0-17 years of age, and 1(1%) targeted maternal-infant dyads in their study. One trial[53] reported enrolling "adults 16 years of age or older", which we categorized as adult participants, despite the discrepancy in definitions. Studies mostly collected and reported sex, rather than gender, however, some studies also reported the responses "prefer not to answer"[54–57], "other"[58,59], "undifferentiated"[54,60], and "unknown"[54] for sex and gender data. Only one trial[61] reported gender as a separate category. Further to this, one trial[52] included two randomizations, and it was unclear how sex was distributed across the two randomizations. Six trials[62–67] reported missing sex data for 16 participants total. Among the trials that clearly reported sex/gender data, 92,277 (54%) female participants and 77,390 (46%) male participants were enrolled in Bayesian RCTs. Notably, fifteen trials focused on female health (i.e., breast cancer, perinatal health) and therefore exclusively enrolled female participants.

### 3.3 Methods for Determining Sample Size

The most common method for SSD in Bayesian efficacy RCTs was a hybrid approach used in 99 (60%) of the 164 trials. In some studies, this hybrid approach was used to evaluate the frequentist operating characteristics of an adaptive design, rather than specifying a single target sample size. Following this, 35 (21%) RCTs used a frequentist approach for SSD and 23 (14%) did not specify how the target sample size was computed. Thus, among the 164 reviewed trials, only 7 (4%) used a fully Bayesian SSD method. These Bayesian SSD calculations were used in RCTs from 2021 (4), 2023 (1), 2024 (1), and 2025 (1), and mostly took place in Europe (4; 57%). They are in immunology,[68,69] oncology,[70] nephrology,[71], infectious diseases,[72] substance abuse, and Alzheimer's disease.[73] Average length criterion was used by one of these trials for SSD, controlling the precision of the posterior distribution through the average length of the posterior credible interval.[71] Five trials averaged over an informative prior distribution to compute expected power and determine the sample size.[68,69,72,74,75] The final SSD is poorly reported but provides a citation to a tutorial that specifies that the maximum sample size should be determined based either on controlling the precision or expected power.[70,76] No trial reported using a utility-based sample size calculation.

The hybrid method for SSD was used in 71% of North American and 71% of international RCTs respectively, compared to 53% of European RCTs. European trials were more likely to use frequentist methods for SSD (37%), and there were minimal differences in the number of RCTs that lacked reporting of SSD across different jurisdictions. A full list of included studies is available in Supplementary Table 3.

**3.4 Outcome Types**

We evaluated the different outcomes used in Bayesian RCTs. The most common outcome type was binary, used by 53 (32%) RCTs. The next most common outcome types were continuous (35; 21%), joint (21; 19%), and ordinal (20; 12%) outcomes. The large number of ordinal outcomes is due to the number of COVID-19 trials that used an ordinal severity score analysed with a proportional odds model. An additional 17 (10%) RCTs used survival outcomes, 8 (5%) used count outcomes, and another 8 (5%) used multiple outcomes. Finally, 1 (1%) RCT used a categorial and continuous outcome, and 1 (1%) did not specify a primary outcome. Of the 141 (86%) RCTs that justified their sample size, 131 (93%) used the same outcome in their SSD and analysis, while 10 (7%) did not.

### 3.5 Prior Distributions

We reviewed the methods used to define the priors for the primary analysis and SSD. Non-informative (65; 40%) or weakly informative priors (44; 27%) were used most frequently. Additionally, 7 (4%) of the RCTs used two co-primary endpoints with one weakly informative and one non-informative prior. Only 23 (14%) RCTs used informative priors, while 17 (10%) RCTs did not specify the prior distribution in the main text or supplementary material. Finally, 8 (5%) RCTs used skeptical priors, which assumed that there was unlikely to be a large effect of the intervention.[77] To classify the priors, the priors with limited information, we primarily used the classification provided by the authors, i.e., "weakly informative" and "vague" priors were classified as weakly informative while "non-informative" priors were classified as such. For papers that did not classify their prior distribution, a normal distribution centered at 0 with a standard deviation of 10 or higher was classified as non-informative as was a beta distribution with parameters less than or equal to (1, 1).

For the RCTs that used an informative prior distribution, 18 (78%) relied on historical data, 3 (13%) used expert opinion, 1 (4%) was based on clinical relevance, and 1 (4%) did not specify the method used to inform the prior.[78] Of the three studies that used expert opinions, two used Bayesian methods for SSD,[68,69] but only one reported using formal expert elicitation.[68] Interestingly, the third study that used an informative prior used a frequentist sample size calculation, meaning that the informative prior was not considered for the SSD. The data-driven priors were either selected from a single previous study (11; 61%) or a meta-analysis (7; 39%). Data-driven priors were either used evenly for the control group alone and both study arms (9; 50%). A range of methods were used to control for prior-data conflict, including down-weighting the prior to represent a small number of patients, power priors, and mixture priors, combining informative and non-informative components.

## 4 Discussion

This review evaluated the current methods for SSD in RCTs with a Bayesian primary efficacy analysis. We found that fully Bayesian methods for SSD are rarely used in practice, with hybrid and frequentist methods used predominantly. Furthermore, no practical Bayesian RCT has used a utility-based SSD. Thus, there is a clear need for improved translation of methods for Bayesian SSD into practice as Bayesian RCTs become more frequent.

Bayesian trials are used in many different diseases areas and are gaining popularity, spurred by the COVID-19 pandemic. Bayesian methods have also analysed a range of different primary outcomes, demonstrating their flexibility. Note that we observed a relatively small number of survival outcomes (17; 10%), potentially as Bayesian survival models require a full likelihood instead of a partial likelihood, common in frequentist survival analysis.[79] Nonetheless, Bayesian methods are being implemented in a range of RCTs, highlighting the importance of improving methods for Bayesian RCTs as they become more popular and familiar.

Throughout this review, we encountered significant challenges identifying the methods for Bayesian SSD as critical elements of the calculation and statistical analysis were often difficult, and at times impossible, to locate. Many elements reported in this review were in the supplementary materials or trial registrations, rather than trial reports. This includes the prior specification, which is critically important to interpreting Bayesian analyses. This limitation may result in inaccuracies in our review but more broadly highlights a critical need for improved Bayesian trial reporting. The CONSORT statement, which provides standards for reporting RCTs, states that the SSD must be included but may need to be extended to account for challenges in Bayesian RCTs.[48,49,80] Specifically, when SSD is based on complex simulations, reporting guidelines are urgently needed. While reporting guidelines for Bayesian analyses exist, (i.e., BayesWatch,[81] ROBUST,[10,11] and BARG,[11]) further guidance specific to SSD is needed.

Other limitations of this project include the restriction to trials published in English, which may have excluded some publications in other languages. Additionally, this review considered only completed studies, potentially omitting unsuccessful trials, or those terminated prematurely without efficacy analysis. Finally, our use of automated screening for our expanded review may have unintentionally excluded some Bayesian RCTs from our review, despite the high performance of our classifier in the test set. While these factors represent limitations, our work constitutes the first comprehensive review in this domain. We hope it will encourage future research aimed at enhancing the quality and consistency of reporting in Bayesian RCTs.

## 5 Conclusion

In conclusion, our review of Bayesian efficacy RCTs demonstrated that Bayesian trials primarily used hybrid or frequentist methods for SSD with only 4% of RCTs using fully Bayesian methods. We also observed generally poor reporting of SSD in Bayesian clinical trials. Thus, better adherence to the CONSORT statement and greater enforcement of the reporting guidelines is needed in Bayesian RCTs.

# 6 Declarations

## 6.1 Ethics approval and consent to participate

Not applicable.

## 6.2 Consent for publication

Not applicable.

## 6.3 Availability of data and materials

All Systematic review data analysed during this study are included in this published article and its supplementary information files.

## 6.4 Competing interests

The authors declare that they have no competing interests.

## 6.5 Funding

AH is funded by Canada Research Chair in Statistical Trial Design; Natural Sciences and Engineering Research Council of Canada (award No. RGPIN-2021-03366). YL and YM are funded by a CIHR award (CIHR Grant #463252). AJ is partially funded by CIHR through the MICYRN RareKids-CAN project. AM and YW are partially funded through a CANSSI-CRT award and AM is partially funded through the CANTRAIN program (CIHR Grant #184898).

### 6.6 Authors contributions

AH led the study design, protocol development, screening, resolving of conflicts, manuscript preparation, and interpretation of findings. YM was involved in the protocol development, and conducted screening, data extraction, quality checks, and analysis, as well as created tables and figures, and prepared the manuscript and supplementary materials. NT supported the study design, screening, and helped develop data extraction forms. AJ conducted screening, data extraction, quality assessment, resolved conflicts, interpreted findings, and edited the manuscript. AM, YL, YW, and LL conducted screening, data extraction and edited the manuscript. OY conducted screening, quality assessment, and resolved conflicts. JC performed the search strategy. All authors critically reviewed the paper and provided edits.

### 6.7 Acknowledgements

Not applicable

# Figures and Tables

## Figure 1. PRISMA Flow Diagram

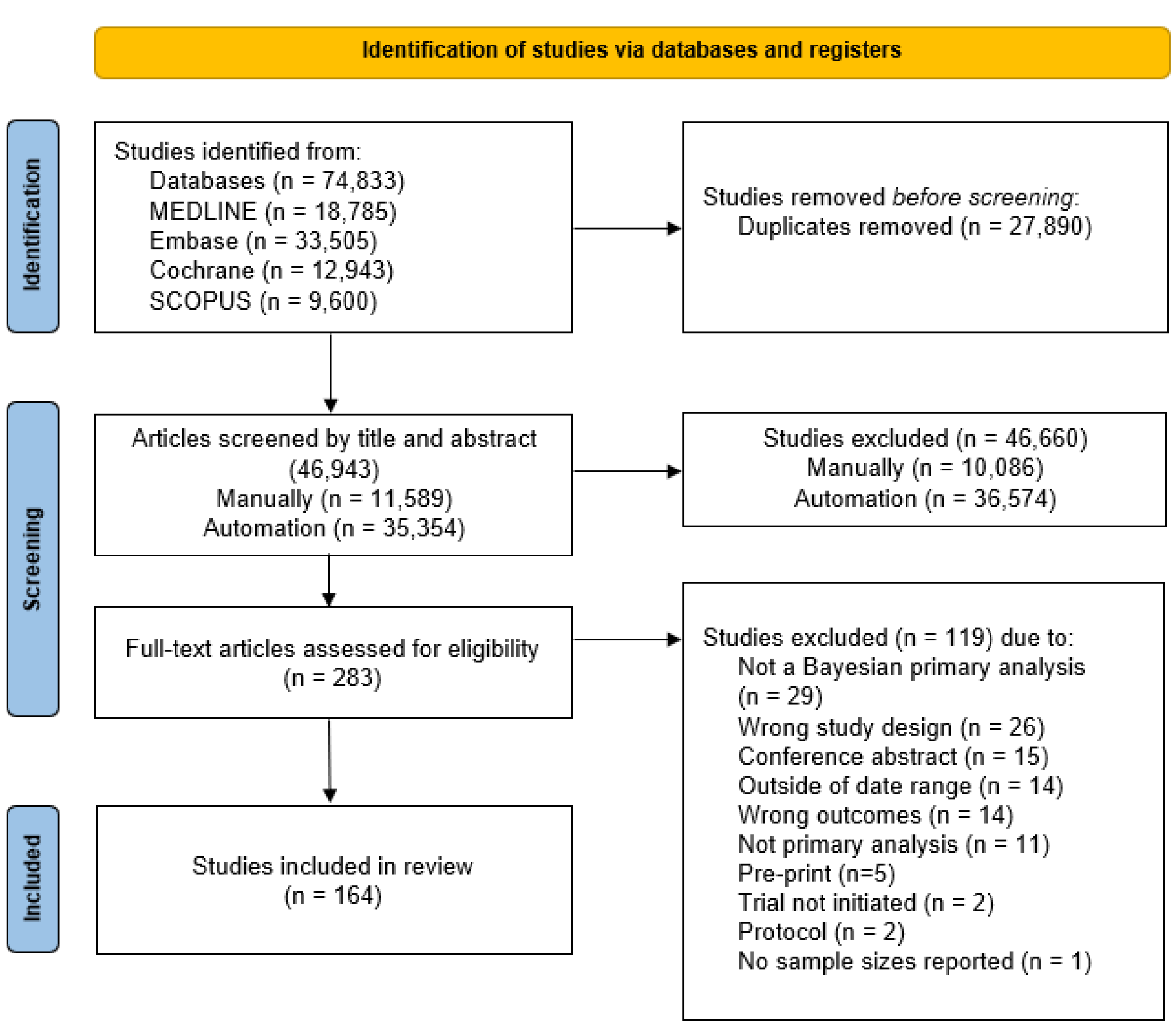

**Figure 2. Number of Bayesian RCTs by Year**

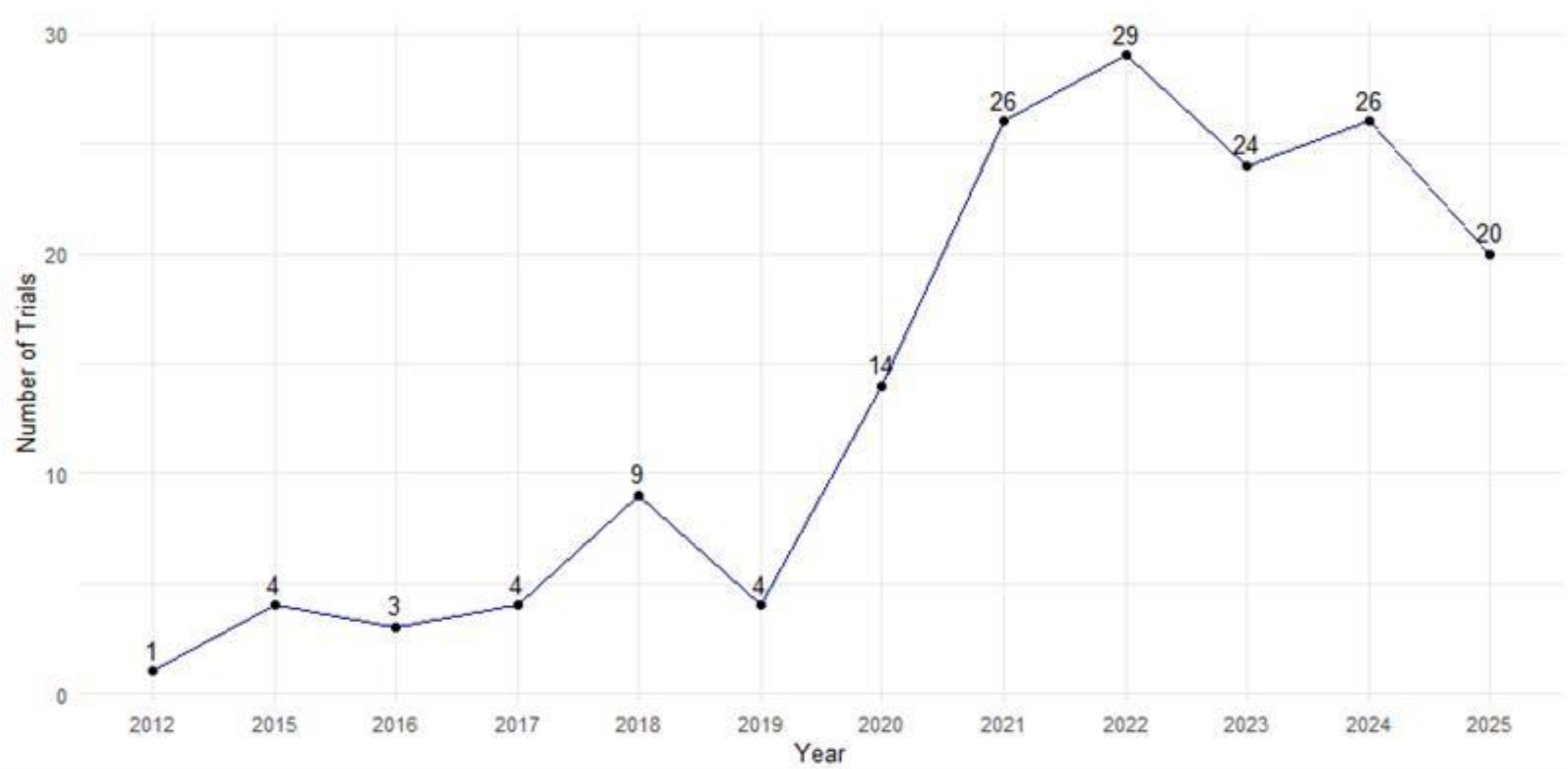

**Table 1. List of disease areas included in the review**

| Disease Area | Number of Trials |
|---|---|
| COVID-19 | 51 |
| Oncology | 16 |
| Neurology | 14 |
| Pulmonology | 9 |
| Cardiology | 9 |
| Infectious Diseases | 7 |
| Psychology | 5 |
| Rheumatology | 5 |
| Substance use | 5 |
| Perinatal Health | 5 |
| Immunology | 4 |
| Obstetrics and Gynecology | 4 |
| Ophthalmology | 3 |
| Endocrinology | 3 |
| Occupational Health | 3 |
| HIV/AIDS | 3 |
| Dermatology | 2 |
| Mental Health | 2 |
| Surgery | 2 |
| Genetic Disorders | 2 |
| Alzheimer's disease | 2 |
| Behavioural | 1 |
| Orthopaedics | 1 |
| Kinesiology | 1 |
| Nephrology | 1 |
| Urology | 1 |
| Somnology | 1 |
| Gastroenterology | 1 |
| Stroke / Neurology | 1 |
| **Number of Trials** | **164** |

# Supplemental Materials

## Supplementary Table 1: Preferred Reporting Items for Systematic reviews and Meta-Analyses (PRISMA) Checklist

| Section and Topic | Item # | Checklist item | Location where item is reported |
|---|---|---|---|
| **TITLE** | | | |
| Title | 1 | Identify the report as a systematic review. | Pg. 1 (Title) |
| **ABSTRACT** | | | |
| Abstract | 2 | See the PRISMA 2020 for Abstracts checklist. | Pg. 2 (Abstract) |
| **INTRODUCTION** | | | |
| Rationale | 3 | Describe the rationale for the review in the context of existing knowledge. | Sec. 1 (Introduction) |
| Objectives | 4 | Provide an explicit statement of the objective(s) or question(s) the review addresses. | Sec. 1 (Introduction) |
| **METHODS** | | | |
| Eligibility criteria | 5 | Specify the inclusion and exclusion criteria for the review and how studies were grouped for the syntheses. | Sec. 2.2.1 |
| Information sources | 6 | Specify all databases, registers, websites, organisations, reference lists and other sources searched or consulted to identify studies. Specify the date when each source was last searched or consulted. | Sec. 8 (References) |
| Search strategy | 7 | Present the full search strategies for all databases, registers and websites, including any filters and limits used. | Supplementary Table 2 |
| Selection process | 8 | Specify the methods used to decide whether a study met the inclusion criteria of the review, including how many reviewers screened each record and each report retrieved, whether they worked independently, and if applicable, details of automation tools used in the process. | Sec. 2.2.3 |
| Data collection process | 9 | Specify the methods used to collect data from reports, including how many reviewers collected data from each report, whether they worked independently, any processes for obtaining or confirming data from study investigators, and if applicable, details of automation tools used in the process. | Sec. 2.2.4 |
| Data items | 10a | List and define all outcomes for which data were sought. Specify whether all results that were compatible with each outcome domain in each study were sought (e.g. for all measures, time points, analyses), and if not, the methods used to decide which results to collect. | Sec. 2.2.4 |
| | 10b | List and define all other variables for which data were sought (e.g. participant and intervention characteristics, funding sources). Describe any assumptions made about any missing or unclear information. | Sec. 2.2.4 |
| Study risk of bias assessment | 11 | Specify the methods used to assess risk of bias in the included studies, including details of the tool(s) used, how many reviewers assessed each study and whether they worked independently, and if applicable, details of automation tools used in the process. | N/A |

| Section and Topic | Item # | Checklist item | Location where item is reported |
|---|---|---|---|
| Effect measures | 12 | Specify for each outcome the effect measure(s) (e.g. risk ratio, mean difference) used in the synthesis or presentation of results. | N/A |
| Synthesis methods | 13a | Describe the processes used to decide which studies were eligible for each synthesis (e.g. tabulating the study intervention characteristics and comparing against the planned groups for each synthesis (item #5)). | N/A |
| | 13b | Describe any methods required to prepare the data for presentation or synthesis, such as handling of missing summary statistics, or data conversions. | N/A |
| | 13c | Describe any methods used to tabulate or visually display results of individual studies and syntheses. | N/A |
| | 13d | Describe any methods used to synthesize results and provide a rationale for the choice(s). If meta-analysis was performed, describe the model(s), method(s) to identify the presence and extent of statistical heterogeneity, and software package(s) used. | N/A |
| | 13e | Describe any methods used to explore possible causes of heterogeneity among study results (e.g. subgroup analysis, meta-regression). | N/A |
| | 13f | Describe any sensitivity analyses conducted to assess robustness of the synthesized results. | N/A |
| Reporting bias assessment | 14 | Describe any methods used to assess risk of bias due to missing results in a synthesis (arising from reporting biases). | N/A |
| Certainty assessment | 15 | Describe any methods used to assess certainty (or confidence) in the body of evidence for an outcome. | N/A |
| **RESULTS** | | | |
| Study selection | 16a | Describe the results of the search and selection process, from the number of records identified in the search to the number of studies included in the review, ideally using a flow diagram. | Sec. 2.2 & Appendix Figure 1 |
| | 16b | Cite studies that might appear to meet the inclusion criteria, but which were excluded, and explain why they were excluded. | Appendix Figure 1 |
| Study characteristics | 17 | Cite each included study and present its characteristics. | Sec. 3.1, 3.2 & Supplementary Table 3 |
| Risk of bias in studies | 18 | Present assessments of risk of bias for each included study. | Sec. 2.2.5 |
| Results of individual studies | 19 | For all outcomes, present, for each study: (a) summary statistics for each group (where appropriate) and (b) an effect estimate and its precision (e.g. confidence/credible interval), ideally using structured tables or plots. | Sec. 3 |
| | 20a | For each synthesis, briefly summarise the characteristics and risk of bias among contributing studies. | Sec. 3 |

| Section and Topic | Item # | Checklist item | Location where item is reported |
|---|---|---|---|
| Results of syntheses | 20b | Present results of all statistical syntheses conducted. If meta-analysis was done, present for each the summary estimate and its precision (e.g. confidence/credible interval) and measures of statistical heterogeneity. If comparing groups, describe the direction of the effect. | Sec. 3 |
| | 20c | Present results of all investigations of possible causes of heterogeneity among study results. | N/A |
| | 20d | Present results of all sensitivity analyses conducted to assess the robustness of the synthesized results. | N/A |
| Reporting biases | 21 | Present assessments of risk of bias due to missing results (arising from reporting biases) for each synthesis assessed. | N/A |
| Certainty of evidence | 22 | Present assessments of certainty (or confidence) in the body of evidence for each outcome assessed. | N/A |
| **DISCUSSION** | | | |
| Discussion | 23a | Provide a general interpretation of the results in the context of other evidence. | Sec. 4 |
| | 23b | Discuss any limitations of the evidence included in the review. | Sec. 4 |
| | 23c | Discuss any limitations of the review processes used. | Sec. 4 |
| | 23d | Discuss implications of the results for practice, policy, and future research. | Sec. 4 & 5 |
| **OTHER INFORMATION** | | | |
| Registration and protocol | 24a | Provide registration information for the review, including register name and registration number, or state that the review was not registered. | Sec. 2.2 |
| | 24b | Indicate where the review protocol can be accessed, or state that a protocol was not prepared. | Sec. 2.2 |
| | 24c | Describe and explain any amendments to information provided at registration or in the protocol. | Sec. 2.2 |
| Support | 25 | Describe sources of financial or non-financial support for the review, and the role of the funders or sponsors in the review. | Sec. 6.5 (Funding) |
| Competing interests | 26 | Declare any competing interests of review authors. | Sec. 6.4 (Competing Interests) |
| Availability of data, code and other materials | 27 | Report which of the following are publicly available and where they can be found: template data collection forms; data extracted from included studies; data used for all analyses; analytic code; any other materials used in the review. | Supplementary Materials (Data) |

# Supplementary Table 2: Detailed Search Strategy

**Literature Search Strategy**

| **Date** | May 16, 2023 |
|---|---|
| **Requester(s)** | Yanara Marks |
| **Topic** | Bayesian sample size calculation methods in clinical trials |
| **Date limit(s)** | December 1, 2009-current |
| **Language** | English |
| **Other limits** | Humans; omitted (books or chapter or conference abstract or conference paper or "conference review" or editorial or erratum or letter or note or short survey or tombstone) in Embase |

| **Database [Platform]** *Initial search run May 16, 2023* | **Results** |
|---|---|
| Ovid MEDLINE(R) and Epub Ahead of Print, In-Process, In-Data-Review & Other Non-Indexed Citations and Daily <1946 to May 15, 2023> | 3538 |
| Embase Classic+Embase <1947 to 2023 Week 19> | 5099 |
| EBM Reviews - Cochrane Central Register of Controlled Trials – April 2023 | 1858 |
| SCOPUS May 16, 2023 | 4098 |
| **TOTAL** | **14593** |
| **Deduplicated total (7043 duplicates removed in Covidence)** | **7550** |

Search strategy:
Set 1 AND Set 2 AND S3

Search methods statement
The search strategy was developed by a professional medical librarian (JC) trained in knowledge synthesis. The search was executed in the following bibliographic databases on May 16, 2023: Ovid MEDLINE(R) and Epub Ahead of Print, In-Process, In-Data-Review & Other Non-Indexed Citations and Daily <1946 to May 15, 2023>; Ovid Embase+Embase Classic (1974–2023 week 19); EBM Reviews Cochrane Central Register of Controlled Trials (April 2023) via Ovid; and Scopus (May 16, 2023) via Elsevier. The search strategy consisted of both controlled vocabulary, such as the National Library of Medicine's MeSH (Medical Subject Headings) and Emtree Subject Headings (Embase), and keywords. Date limits of December 2009-2023 were applied. English language and human study type filters were applied. Additionally in Embase and Scopus, conference abstracts and other particular publication types were omitted. Results were uploaded to Covidence, where duplicates were automatically removed. See Appendix [To be added] for the detailed search strategy.

Ovid MEDLINE(R) and Epub Ahead of Print, In-Process, In-Data-Review & Other Non-Indexed Citations and Daily <1946 to May 15, 2023>

| # | Searches | Results |
|---|---|---|
| 1 | "Clinical Trials, Phase III as Topic"/ or " Clinical Trials, Phase IV as Topic "/ or ((Clinical trial* or drug evaluation* or evaluation stud* or efficacy) adj3 ("phase 3" or "phase III" or "phase 4" or "phase IV")).tw,kw,kf. | 20321 |

| 2 | ("Clinical Trial, Phase III" or "Clinical Trial, Phase IV").pt. | 23963 |
|---|---|---|
| 3 | ((platform* or complet* or adaptive*) and trial*).tw,kf. | 191143 |
| 4 | (("phase 3" or "phase III" or "phase 4" or "phase IV") adj3 (study or studies or trial*)).ti,ab,hw,kf. | 63342 |
| 5 | or/1-4 | 245205 |
| 6 | "Clinical Trials, Phase I as Topic"/ or ((Clinical trial* or drug evaluation* or evaluation stud* or efficacy) adj3 ("phase 1" or "phase I")).tw,kw,kf. | 15262 |
| 7 | "Clinical Trial, Phase I".pt. | 24899 |
| 8 | or/6-7 | 37496 |
| 9 | 5 not 8 | 238034 |
| 10 | Bayes Theorem/ or ((Theorem* or theory* or forecast* or analy* or approach* or method$ or estimat* or predict* or calculat* or statistical model$ or analysis model* or study design* or design) adj4 (Bayes or Bayesian*)).ti,ab,kf. | 65072 |
| 11 | ("Posterior probability" or "Posterior distribution" or "Prior distribution" or "Uninformative prior" or prior or "Credible interval" or "Hierarchical model*" or "Dynamic borrowing" or "Predictive probability").ti,ab,kf. | 709277 |
| 12 | or/10-11 | 763147 |
| 13 | Sample Size/ or ("sample size*" or "adaptive-sample-size").ti,ab,kf. | 106860 |
| 14 | (sampl* adj2 (size$ or randomi* or proportion* or stratified or systematic or cluster* or representative or group$ or population$ or select* or calculat*)).ti,ab,kf. | 249278 |
| 15 | (patient$ adj4 (randomly assign* or baseline*)).ti,ab,kf. | 79236 |
| 16 | (patient$ adj9 (random* or cohort$)).ti,ab,kf. | 544918 |
| 17 | (Subgroup$ or "control group$").ti,ab,kf. | 844049 |
| 18 | or/13-17 | 1579159 |
| 19 | 9 and 12 and 18 | 5373 |
| 20 | limit 19 to dt=20091201-20231231 | 4231 |
| 21 | limit 19 to ez=20091201-20231231 | 4225 |
| 22 | limit 19 to ed=20091201-20231231 | 3713 |
| 23 | or/20-22 | 4289 |
| 24 | limit 23 to (english language and humans) | 3553 |
| 25 | remove duplicates from 24 | 3538 |

Embase Classic+Embase <1947 to 2023 Week 19>

| # | Searches | Results |
|---|---|---|
| 1 | "phase 3 clinical trial (topic)"/ or "phase 4 Clinical Trial (topic)"/ or ((Clinical trial* or drug evaluation* or evaluation stud* or efficacy) adj3 ("phase 3" or "phase III" or "phase 4" or "phase IV")).tw,kw,kf. | 66290 |

| 2 | ((platform* or complet* or adaptive*) and trial*).tw,kf. | 321340 |
|---|---|---|
| 3 | (("phase 3" or "phase III" or "phase 4" or "phase IV") adj3 (study or studies or trial*)).ti,ab,hw,kf. | 164619 |
| 4 | or/1-3 | 462533 |
| 5 | "Phase 1 Clinical Trial (Topic)"/ or ((Clinical trial* or drug evaluation* or evaluation stud* or efficacy) adj3 ("phase 1" or "phase I")).tw,kw,kf. | 47850 |
| 6 | 4 not 5 | 442023 |
| 7 | Bayes Theorem/ or ((Theorem* or theory* or forecast* or analy* or approach* or method$ or estimat* or predict* or calculat* or statistical model$ or analysis model* or study design* or design) adj4 (Bayes or Bayesian*)).ti,ab,kf. | 67380 |
| 8 | ("Posterior probability" or "Posterior distribution" or "Prior distribution" or "Uninformative prior" or prior or "Credible interval" or "Hierarchical model*" or "Dynamic borrowing" or "Predictive probability").ti,ab,kf. | 1237685 |
| 9 | or/7-8 | 1292792 |
| 10 | Sample Size/ or ("sample size*" or "adaptive-sample-size").ti,ab,kf. | 197339 |
| 11 | (sampl* adj2 (size$ or randomi* or proportion* or stratified or systematic or cluster* or representative or group$ or population$ or select* or calculat*)).ti,ab,kf. | 361901 |
| 12 | (patient$ adj4 ("randomly assign*" or baseline*)).ti,ab,kf. | 158241 |
| 13 | (patient$ adj9 (random* or cohort$)).ti,ab,kf. | 981728 |
| 14 | (Subgroup$ or "control group$").ti,ab,kf. | 1259310 |
| 15 | or/10-14 | 2535484 |
| 16 | 6 and 9 and 15 | 19301 |
| 17 | limit 16 to dd=20091201-20231231 | 14694 |
| 18 | limit 16 to dc=20091201-20231231 | 18024 |
| 19 | 17 or 18 | 18949 |
| 20 | limit 19 to (human and english language) | 17871 |
| 21 | limit 20 to (books or chapter or conference abstract or conference paper or "conference review" or editorial or erratum or letter or note or short survey or tombstone) | 12430 |
| 22 | 20 not 21 | 5441 |
| 23 | remove duplicates from 22 | 5099 |

EBM Reviews - Cochrane Central Register of Controlled Trials – April 2023

| # | Searches | Results |
|---|---|---|
| 1 | "Clinical Trials, Phase III as Topic"/ or " Clinical Trials, Phase IV as Topic "/ or ((Clinical trial* or drug evaluation* or evaluation stud* or efficacy) adj3 ("phase 3" or "phase III" or "phase 4" or "phase IV")).tw,kw,kf. | 41049 |
| 2 | ((platform* or complet* or adaptive*) and trial*).tw,kf. | 152632 |
| 3 | (("phase 3" or "phase III" or "phase 4" or "phase IV") adj3 (study or studies or trial*)).ti,ab,hw,kf. | 70194 |
| 4 | or/1-3 | 211032 |
| 5 | "Clinical Trials, Phase I as Topic"/ or ((Clinical trial* or drug evaluation* or evaluation stud* or efficacy) adj3 ("phase 1" or "phase I")).tw,kw,kf. | 12733 |
| 6 | 4 not 5 | 207023 |
| 7 | Bayes Theorem/ or ((Theorem* or theory* or forecast* or analy* or approach* or method$ or estimat* or predict* or calculat* or statistical model$ or analysis model* or study design* or design) adj4 (Bayes or Bayesian*)).ti,ab,kf. | 2309 |
| 8 | ("Posterior probability" or "Posterior distribution" or "Prior distribution" or "Uninformative prior" or prior or "Credible interval" or "Hierarchical model*" or "Dynamic borrowing" or "Predictive probability").ti,ab,kf. | 128277 |
| 9 | or/7-8 | 129846 |
| 10 | Sample Size/ or ("sample size*" or "adaptive-sample-size").ti,ab,kf. | 33279 |
| 11 | (sampl* adj2 (size$ or randomi* or proportion* or stratified or systematic or cluster* or representative or group$ or population$ or select* or calculat*)).ti,ab,kf. | 41459 |
| 12 | (patient$ adj4 (randomly assign* or baseline*)).ti,ab,kf. | 64439 |
| 13 | (patient$ adj9 (random* or cohort$)).ti,ab,kf. | 403018 |
| 14 | (Subgroup$ or "control group$").ti,ab,kf. | 294441 |
| 15 | or/10-14 | 647170 |
| 16 | 6 and 9 and 15 | 14110 |
| 17 | limit 16 to english language | 14063 |
| 18 | limit 17 to yr="2010 -Current" | 12659 |
| 19 | limit 18 to (clinical trial protocol or comment or conference proceeding or corrected publication or dissertation thesis or editorial or erratum or "expression of concern" or letter or multimedia or note or report or retraction or trial registry record) | 10722 |
| 20 | 18 not 19 | 1937 |
| 21 | Animals/ or animal*.mp. | 37765 |
| 22 | 20 not 21 | 1915 |
| 23 | remove duplicates from 22 | 1858 |

SCOPUS
May 16, 2023
n=4098

( ( TITLE-ABS-KEY ( "sample size*" OR "adaptive-sample-size" ) ) OR ( TITLE-ABS-KEY ( sampl* W/2 ( size? OR randomi* OR proportion* OR stratified OR systematic OR cluster* OR representative OR group? OR population? OR select* OR calculat* ) ) ) OR ( TITLE-ABS-KEY ( patient? W/4 ( "randomly assign*" OR baseline* ) ) ) OR ( TITLE-ABS-KEY ( patient? W/9 ( random* OR cohort? ) ) ) OR ( TITLE-ABS-KEY ( subgroup? OR "control group?" ) ) ) AND ( ( TITLE-ABS-KEY ( ( theorem* OR theory* OR forecast* OR analy* OR approach* OR method? OR estimat* OR predict* OR calculat* OR "statistical model?" OR "analysis model*" OR "study design*" OR design ) W/4 ( bayes OR bayesian* ) ) ) OR ( TITLE-ABS-KEY ( "Posterior probability" OR "Posterior distribution" OR "Prior distribution" OR "Uninformative prior" OR prior OR "Credible interval" OR "Hierarchical model*" OR "Dynamic borrowing" OR "Predictive probability" ) ) ) AND ( ( ( TITLE-ABS-KEY ( ( "Clinical trial*" OR "drug evaluation*" OR "evaluation stud*" OR efficacy ) W/3 ( "phase 3" OR "phase III" OR "phase 4" OR "phase IV" ) ) ) OR ( TITLE-ABS-KEY ( ( platform* OR complet* OR adaptive* ) AND trial* ) ) OR ( ( ( "phase 3" OR "phase III" OR "phase 4" OR "phase IV" ) W/3 ( study OR studies OR trial* ) ) ) ) AND NOT ( TITLE-ABS-KEY ( ( "Clinical trial*" OR "drug evaluation*" OR "evaluation stud*" OR efficacy ) W/3 ( "phase 1" OR "phase I" ) ) ) ) AND PUBYEAR > 2009 AND PUBYEAR < 2024 AND ( LIMIT-TO ( DOCTYPE , "ar" ) ) AND ( LIMIT-TO ( LANGUAGE , "English" ) ) AND ( LIMIT-TO ( EXACTKEYWORD , "Humans" ) ) AND ( LIMIT-TO ( PUBSTAGE , "final" ) ) AND ( LIMIT-TO ( SRCTYPE , "j" ) )

**Literature Search Strategy - Updated**

| **Date** | January 16, 2025 - update |
|---|---|
| **Requester(s)** | Yanara Marks |
| **Topic** | Bayesian sample size calculation methods in clinical trials |
| **Date limit(s)** | May 16 2023-Current |
| **Language** | English |
| **Other limits** | Humans; omitted (books or chapter or conference abstract or conference paper or "conference review" or editorial or erratum or letter or note or short survey or tombstone) in Embase |

| **Database [Platform]** *Search updated January 16, 2025* | **Results** |
|---|---|
| Ovid MEDLINE(R) and Epub Ahead of Print, In-Process, In-Data-Review & Other Non-Indexed Citations and Daily <1946 to January 15, 2025> | 668 |
| Ovid Embase Classic+Embase <1947 to 2025 Week 02> | 934 |
| Ovid EBM Reviews - Cochrane Central Register of Controlled Trials – December 2024 | 2485 |
| Elsevier SCOPUS January 16, 2025 | 502 |
| **TOTAL** | **4,589** |
| **Deduplicated total (1,531 duplicates removed in Covidence)** | **3,058** |

**Search methods statement**

The search strategy was developed by a professional medical librarian (JC) trained in knowledge synthesis. The search was executed in the following bibliographic databases on May 16, 2023: Ovid MEDLINE(R) and Epub Ahead of Print, In-Process, In-Data-Review & Other Non-Indexed Citations and Daily <1946 to May 15, 2023>; Ovid Embase+Embase Classic (1974–2023 week 19); EBM Reviews Cochrane Central Register of Controlled Trials (April 2023) via Ovid; and Scopus (May 16, 2023) via Elsevier. The search strategy consisted of both controlled vocabulary, such as the National Library of Medicine's MeSH (Medical Subject Headings) and Emtree Subject Headings (Embase), and keywords. Date limits of December 2009-2023 were applied. English language and human study type filters were applied. Additionally in Embase and Scopus, conference abstracts and other particular publication types were omitted. Results were uploaded to Covidence, where duplicates were automatically removed. The search was updated January 16, 2025 using the same parameters, except the date range.

Ovid MEDLINE(R) and Epub Ahead of Print, In-Process, In-Data-Review & Other Non-Indexed Citations and Daily <1946 to January 15, 2025>

| # | Searches | Results |
|---|---|---|
| 1 | "Clinical Trials, Phase III as Topic"/ or " Clinical Trials, Phase IV as Topic "/ or ((Clinical trial* or drug evaluation* or evaluation stud* or efficacy) adj3 ("phase 3" or "phase III" or "phase 4" or "phase IV")).tw,kw,kf. | 22336 |
| 2 | ("Clinical Trial, Phase III" or "Clinical Trial, Phase IV").pt. | 26356 |
| 3 | ((platform* or complet* or adaptive*) and trial*).tw,kf. | 218073 |
| 4 | (("phase 3" or "phase III" or "phase 4" or "phase IV") adj3 (study or studies or trial*)).ti,ab,hw,kf. | 70218 |
| 5 | or/1-4 | 277796 |
| 6 | "Clinical Trials, Phase I as Topic"/ or ((Clinical trial* or drug evaluation* or evaluation stud* or efficacy) adj3 ("phase 1" or "phase I")).tw,kw,kf. | 16423 |
| 7 | "Clinical Trial, Phase I".pt. | 26877 |
| 8 | or/6-7 | 40381 |
| 9 | 5 not 8 | 269950 |
| 10 | Bayes Theorem/ or ((Theorem* or theory* or forecast* or analy* or approach* or method$ or estimat* or predict* or calculat* or statistical model$ or analysis model* or study design* or design) adj4 (Bayes or Bayesian*)).ti,ab,kf. | 74827 |
| 11 | ("Posterior probability" or "Posterior distribution" or "Prior distribution" or "Uninformative prior" or prior or "Credible interval" or "Hierarchical model*" or "Dynamic borrowing" or "Predictive probability").ti,ab,kf. | 780030 |
| 12 | or/10-11 | 841956 |
| 13 | Sample Size/ or ("sample size*" or "adaptive-sample-size").ti,ab,kf. | 125675 |
| 14 | (sampl* adj2 (size$ or randomi* or proportion* or stratified or systematic or cluster* or representative or group$ or population$ or select* or calculat*)).ti,ab,kf. | 286626 |
| 15 | (patient$ adj4 (randomly assign* or baseline*)).ti,ab,kf. | 89264 |
| 16 | (patient$ adj9 (random* or cohort$)).ti,ab,kf. | 628341 |
| 17 | (Subgroup$ or "control group$").ti,ab,kf. | 951424 |
| 18 | or/13-17 | 1793838 |
| 19 | 9 and 12 and 18 | 6227 |
| 20 | limit 19 to dt=20230516-20250116 | 862 |
| 21 | limit 19 to ed=20230516-20250116 | 688 |
| 22 | limit 19 to ez=20230516-20250116 | 857 |
| 23 | or/20-22 | 939 |
| 24 | limit 23 to (english language and humans) | 675 |
| 25 | remove duplicates from 24 | 668 |

Ovid Embase Classic+Embase <1947 to 2025 Week 02>

| # | Searches | Results |
|---|---|---|
| 1 | "phase 3 clinical trial (topic)"/ or "phase 4 Clinical Trial (topic)"/ or ((Clinical trial* or drug evaluation* or evaluation stud* or efficacy) adj3 ("phase 3" or "phase III" or "phase 4" or "phase IV")).tw,kw,kf. | 72430 |
| 2 | ((platform* or complet* or adaptive*) and trial*).tw,kf. | 357895 |
| 3 | (("phase 3" or "phase III" or "phase 4" or "phase IV") adj3 (study or studies or trial*)).ti,ab,hw,kf. | 185220 |
| 4 | or/1-3 | 516002 |
| 5 | "Phase 1 Clinical Trial (Topic)"/ or ((Clinical trial* or drug evaluation* or evaluation stud* or efficacy) adj3 ("phase 1" or "phase I")).tw,kw,kf. | 52199 |
| 6 | 4 not 5 | 493427 |
| 7 | Bayes Theorem/ or ((Theorem* or theory* or forecast* or analy* or approach* or method$ or estimat* or predict* or calculat* or statistical model$ or analysis model* or study design* or design) adj4 (Bayes or Bayesian*)).ti,ab,kf. | 75150 |
| 8 | ("Posterior probability" or "Posterior distribution" or "Prior distribution" or "Uninformative prior" or prior or "Credible interval" or "Hierarchical model*" or "Dynamic borrowing" or "Predictive probability").ti,ab,kf. | 1348931 |
| 9 | 7 or 8 | 1410413 |
| 10 | Sample Size/ or ("sample size*" or "adaptive-sample-size").ti,ab,kf. | 229319 |
| 11 | (sampl* adj2 (size$ or randomi* or proportion* or stratified or systematic or cluster* or representative or group$ or population$ or select* or calculat*)).ti,ab,kf. | 405839 |
| 12 | (patient$ adj4 ("randomly assign*" or baseline*)).ti,ab,kf. | 174284 |
| 13 | (patient$ adj9 (random* or cohort$)).ti,ab,kf. | 1094859 |
| 14 | (Subgroup$ or "control group$").ti,ab,kf. | 1380879 |
| 15 | or/10-14 | 2803893 |
| 16 | 6 and 9 and 15 | 22519 |
| 17 | limit 16 to dc=20230516-20250116 | 3934 |
| 18 | limit 17 to (human and english language) | 3898 |
| 19 | limit 18 to (books or chapter or conference abstract or conference paper or "conference review" or editorial or erratum or letter or note or short survey or tombstone) | 2955 |
| 20 | 18 not 19 | 943 |
| 21 | remove duplicates from 20 | 934 |

Ovid EBM Reviews - Cochrane Central Register of Controlled Trials – December 2024

| # | Searches | Results |
|---|---|---|
| 1 | "Clinical Trials, Phase III as Topic"/ or " Clinical Trials, Phase IV as Topic "/ or ((Clinical trial* or drug evaluation* or evaluation stud* or efficacy) adj3 ("phase 3" or "phase III" or "phase 4" or "phase IV")).tw,kw,kf. | 51428 |

| 2 | ("Clinical Trial, Phase III" or "Clinical Trial, Phase IV").pt. | 0 |
|---|---|---|
| 3 | ((platform* or complet* or adaptive*) and trial*).tw,kf. | 181525 |
| 4 | (("phase 3" or "phase III" or "phase 4" or "phase IV") adj3 (study or studies or trial*)).ti,ab,hw,kf. | 82397 |
| 5 | or/1-4 | 249618 |
| 6 | "Clinical Trials, Phase I as Topic"/ or ((Clinical trial* or drug evaluation* or evaluation stud* or efficacy) adj3 ("phase 1" or "phase I")).tw,kw,kf. | 14996 |
| 7 | "Clinical Trial, Phase I".pt. | 0 |
| 8 | or/6-7 | 14996 |
| 9 | 5 not 8 | 244893 |
| 10 | Bayes Theorem/ or ((Theorem* or theory* or forecast* or analy* or approach* or method$ or estimat* or predict* or calculat* or statistical model$ or analysis model* or study design* or design) adj4 (Bayes or Bayesian*)).ti,ab,kf. | 2999 |
| 11 | ("Posterior probability" or "Posterior distribution" or "Prior distribution" or "Uninformative prior" or prior or "Credible interval" or "Hierarchical model*" or "Dynamic borrowing" or "Predictive probability").ti,ab,kf. | 147317 |
| 12 | or/10-11 | 149314 |
| 13 | Sample Size/ or ("sample size*" or "adaptive-sample-size").ti,ab,kf. | 39800 |
| 14 | (sampl* adj2 (size$ or randomi* or proportion* or stratified or systematic or cluster* or representative or group$ or population$ or select* or calculat*)).ti,ab,kf. | 49491 |
| 15 | (patient$ adj4 (randomly assign* or baseline*)).ti,ab,kf. | 73608 |
| 16 | (patient$ adj9 (random* or cohort$)).ti,ab,kf. | 459028 |
| 17 | (Subgroup$ or "control group$").ti,ab,kf. | 351299 |
| 18 | or/13-17 | 750672 |
| 19 | 9 and 12 and 18 | 16764 |
| 20 | limit 19 to (yr="2023 -Current" and english language) | 2531 |
| 21 | remove duplicates from 20 | 2485 |

Elsevier SCOPUS
January 16, 2025
n=502

( ( TITLE-ABS-KEY ( "sample size*" OR "adaptive-sample-size" ) ) OR ( TITLE-ABS-KEY ( sampl* W/2 ( size? OR randomi* OR proportion* OR stratified OR systematic OR cluster* OR representative OR group? OR population? OR select* OR calculat* ) ) ) OR ( TITLE-ABS-KEY ( patient? W/4 ( "randomly assign*" OR baseline* ) ) ) OR ( TITLE-ABS-KEY ( patient? W/9 ( random* OR cohort? ) ) ) OR ( TITLE-ABS-KEY ( subgroup? OR "control group?" ) ) ) AND ( ( TITLE-ABS-KEY ( ( theorem* OR theory* OR forecast* OR analy* OR approach* OR method? OR estimat* OR predict* OR calculat* OR "statistical model?" OR "analysis model*" OR "study design*" OR design ) W/4 ( bayes OR bayesian* ) ) ) OR ( TITLE-ABS-KEY ( "Posterior probability" OR "Posterior distribution" OR "Prior distribution" OR "Uninformative prior" OR prior OR "Credible interval" OR "Hierarchical model*" OR "Dynamic borrowing" OR "Predictive probability" ) ) ) AND ( ( ( TITLE-ABS-KEY ( ( "Clinical trial*" OR "drug evaluation*" OR "evaluation stud*" OR efficacy ) W/3 ( "phase 3" OR "phase III" OR "phase 4" OR "phase IV" ) ) ) OR ( TITLE-ABS-KEY ( ( platform* OR complet* OR adaptive* ) AND trial* ) ) OR ( ( ( "phase 3" OR "phase III" OR "phase 4" OR "phase IV" ) W/3 ( study OR studies OR trial* ) ) ) ) ) AND NOT ( TITLE-ABS-KEY ( ( "Clinical trial*" OR "drug evaluation*" OR "evaluation stud*" OR efficacy ) W/3 ( "phase 1" OR "phase I" ) ) ) ) ) AND PUBYEAR > 2023 AND PUBYEAR < 2025 AND ( LIMIT-TO ( DOCTYPE , "ar" ) ) AND ( LIMIT-TO ( LANGUAGE , "English" ) ) AND ( LIMIT-TO ( EXACTKEYWORD , "Humans" ) ) AND ( LIMIT-TO ( PUBSTAGE , "final" ) ) AND ( LIMIT-TO ( SRCTYPE , "j" ) )

**Literature Search Strategy - Updated**

| **Date** | July 22, 2025 |
|---|---|
| **Requester(s)** | Yanara Marks, Anna Heath |
| **Topic** | Bayesian sample size calculation methods in clinical trials |
| **Date limit(s)** | May 16 2023-Current |
| **Language** | English |
| **Other limits** | Humans; omitted (books or chapter or conference abstract or conference paper or "conference review" or editorial or erratum or letter or note or short survey or tombstone) in Embase |

| **Database [Platform]** *Search updated July 22, 2025* | **Results** |
|---|---|
| Ovid MEDLINE(R) and Epub Ahead of Print, In-Process, In-Data-Review & Other Non-Indexed Citations and Daily <1946 to July 21, 2025> | 14579 |
| Ovid Embase Classic+Embase <1947 to 2025 Week 29> | 27472 |
| Ovid EBM Reviews - Cochrane Central Register of Controlled Trials – June 2025 | 8600 |
| Elsevier SCOPUS July 22, 2025 (only 5000 exported due to SCOPUS interface limitations) | 5000 |
| **TOTAL** | **55,651** |
| **Deduplicated total (18,956 duplicates removed in Covidence)** | **36,695** |

Search methods statement

Ovid MEDLINE(R) and Epub Ahead of Print, In-Process, In-Data-Review & Other Non-Indexed Citations and Daily <1946 to July 21, 2025>

| # | **Searches** | **Results** |
|---|---|---|
| 1 | "Clinical Trials, Phase III as Topic"/ or " Clinical Trials, Phase IV as Topic "/ or ((Clinical trial* or drug evaluation* or evaluation stud* or efficacy) adj3 ("phase 3" or "phase III" or "phase 4" or "phase IV")).tw,kw,kf. | 23152 |
| 2 | ("Clinical Trial, Phase III" or "Clinical Trial, Phase IV").pt. | 27720 |
| 3 | ((platform* or complet* or adaptive*) and trial*).tw,kf. | 228195 |
| 4 | (("phase 3" or "phase III" or "phase 4" or "phase IV") adj3 (study or studies or trial*)).ti,ab,hw,kf. | 72984 |
| 5 | or/1-4 | 290202 |
| 6 | "Clinical Trials, Phase I as Topic"/ or ((Clinical trial* or drug evaluation* or evaluation stud* or efficacy) adj3 ("phase 1" or "phase I")).tw,kw,kf. | 16835 |

| 7 | "Clinical Trial, Phase I".pt. | 27784 |
|---|---|---|
| 8 | 6 or 7 | 41574 |
| 9 | 5 not 8 | 282052 |
| 10 | Bayes Theorem/ or ((Theorem* or theory* or forecast* or analy* or approach* or method$ or estimat* or predict* or calculat* or statistical model$ or analysis model* or study design* or design) adj4 (Bayes or Bayesian*)).ti,ab,kf. | 77996 |
| 11 | ("Posterior probability" or "Posterior distribution" or "Prior distribution" or "Uninformative prior" or prior or "Credible interval" or "Hierarchical model*" or "Dynamic borrowing" or "Predictive probability").ti,ab,kf. | 804644 |
| 12 | 10 or 11 | 869243 |
| 13 | 9 and 12 | 20861 |
| 14 | (Animals/ or Animal Experimentation/ or "Models, Animal"/ or (animal* or nonhuman or non human or Bovine or Cow? or Monkey? or rat or rats or mouse or mice or rabbit or rabbit or pig or pigs or porcine or dog or dogs or hamster or hamsters or fish or chicken or chickens or sheep or cat or cats or raccoon or raccoons or rodent* or horse or horses or racehorse or racehorses or beagle*).ti,ab.) not (Humans/ or (human* or participant* or patient or patients).ti,ab.) | 5338192 |
| 15 | 13 not 14 | 20388 |
| 16 | limit 15 to (address or autobiography or bibliography or biography or case reports or classical article or clinical conference or clinical trial, phase i or collected work or comment or comparative study or congress or consensus development conference or consensus development conference, nih or "corrected and republished article" or dataset or dictionary or directory or duplicate publication or editorial or english abstract or electronic supplementary materials or "expression of concern" or festschrift or government publication or guideline or historical article or interactive tutorial or interview or introductory journal article or lecture or legal case or legislation or letter or news or newspaper article or observational study, veterinary or observational study or overall or patient education handout or periodical index or personal narrative or portrait or practice guideline or published erratum or randomized controlled trial, veterinary or retracted publication or retraction notice or "scientific integrity review" or "scoping review" or study guide or technical report or video-audio media or webcast) | 2436 |
| 17 | 15 not 16 | 17952 |
| 18 | limit 17 to (english language and yr="2009 -Current") | 14579 |

Ovid Embase Classic+Embase <1947 to 2025 Week 29>

| # | Searches | Results |
|---|---|---|
| 1 | "phase 3 clinical trial (topic)"/ or "phase 4 Clinical Trial (topic)"/ or ((Clinical trial* or drug evaluation* or evaluation stud* or efficacy) adj3 ("phase 3" or "phase III" or "phase 4" or "phase IV")).tw,kw,kf. | 77537 |
| 2 | ((platform* or complet* or adaptive*) and trial*).tw,kf. | 429548 |
| 3 | (("phase 3" or "phase III" or "phase 4" or "phase IV") adj3 (study or studies or trial*)).ti,ab,hw,kf. | 227723 |
| 4 | or/1-3 | 624242 |
| 5 | "Phase 1 Clinical Trial (Topic)"/ or ((Clinical trial* or drug evaluation* or evaluation stud* or efficacy) adj3 ("phase 1" or "phase I")).tw,kw,kf. | 56830 |
| 6 | 4 not 5 | 600290 |
| 7 | Bayes Theorem/ or ((Theorem* or theory* or forecast* or analy* or approach* or method$ or estimat* or predict* or calculat* or statistical model$ or analysis model* or study design* or design) adj4 (Bayes or Bayesian*)).ti,ab,kf. | 79320 |
| 8 | ("Posterior probability" or "Posterior distribution" or "Prior distribution" or "Uninformative prior" or prior or "Credible interval" or "Hierarchical model*" or "Dynamic borrowing" or "Predictive probability").ti,ab,kf. | 1455886 |
| 9 | or/7-8 | 1520718 |
| 10 | 6 and 9 | 79446 |
| 11 | (Animal/ or Animal Experiment/ or "Animal Model"/ or (animal* or nonhuman or non human or Bovine or Cow? or Monkey? or rat or rats or mouse or mice or rabbit or rabbit or pig or pigs or porcine or dog or dogs or hamster or hamsters or fish or chicken or chickens or sheep or cat or cats or raccoon or raccoons or rodent* or horse or horses or racehorse or racehorses or beagle*).ti,ab.) not (Human/ or (human* or participant* or patient or patients).ti,ab.) | 5706705 |
| 12 | 10 not 11 | 78823 |
| 13 | limit 12 to (books or chapter or conference abstract or conference paper or "conference review" or data paper or editorial or erratum or letter or note or "preprint (unpublished, non-peer reviewed)" or "review" or short survey or tombstone) | 44941 |
| 14 | 12 not 13 | 33882 |
| 15 | limit 14 to (english language and yr="2009 -Current") | 27472 |

Ovid EBM Reviews - Cochrane Central Register of Controlled Trials June 2025

| # | Searches | Results |
|---|---|---|
| 1 | "Clinical Trials, Phase III as Topic"/ or " Clinical Trials, Phase IV as Topic "/ or ((Clinical trial* or drug evaluation* or evaluation stud* or efficacy) adj3 ("phase 3" or "phase III" or "phase 4" or "phase IV")).tw,kw,kf. | 51913 |
| 2 | ("Clinical Trial, Phase III" or "Clinical Trial, Phase IV").pt. | 0 |
| 3 | ((platform* or complet* or adaptive*) and trial*).tw,kf. | 183469 |
| 4 | (("phase 3" or "phase III" or "phase 4" or "phase IV") adj3 (study or studies or trial*)).ti,ab,hw,kf. | 82766 |
| 5 | or/1-4 | 251772 |

| 6 | "Clinical Trials, Phase I as Topic"/ or ((Clinical trial* or drug evaluation* or evaluation stud* or efficacy) adj3 ("phase 1" or "phase I")).tw,kw,kf. | 15026 |
|---|---|---|
| 7 | "Clinical Trial, Phase I".pt. | 0 |
| 8 | 6 or 7 | 15026 |
| 9 | 5 not 8 | 246985 |
| 10 | Bayes Theorem/ or ((Theorem* or theory* or forecast* or analy* or approach* or method$ or estimat* or predict* or calculat* or statistical model$ or analysis model* or study design* or design) adj4 (Bayes or Bayesian*)).ti,ab,kf. | 3075 |
| 11 | ("Posterior probability" or "Posterior distribution" or "Prior distribution" or "Uninformative prior" or prior or "Credible interval" or "Hierarchical model*" or "Dynamic borrowing" or "Predictive probability").ti,ab,kf. | 149481 |
| 12 | 10 or 11 | 151524 |
| 13 | 9 and 12 | 34239 |
| 14 | (Animals/ or Animal Experimentation/ or "Models, Animal"/ or (animal* or nonhuman or non human or Bovine or Cow? or Monkey? or rat or rats or mouse or mice or rabbit or rabbit or pig or pigs or porcine or dog or dogs or hamster or hamsters or fish or chicken or chickens or sheep or cat or cats or raccoon or raccoons or rodent* or horse or horses or racehorse or racehorses or beagle*).ti,ab.) not (Humans/ or (human* or participant* or patient or patients).ti,ab.) | 10050 |
| 15 | limit 14 to (yr="2009 -Current" and english language) | 8600 |

SCOPUS
July 22, 2025
n=27,291 (only 5000 exported due to SCOPUS interface limitations)

( ( ( TITLE-ABS-KEY ( ( theorem* OR theory* OR forecast* OR analy* OR approach* OR method? OR estimat* OR predict* OR calculat* OR "statistical model?" OR "analysis model*" OR "study design*" OR design ) W/4 ( bayes OR bayesian* ) ) ) OR ( TITLE-ABS-KEY ( "Posterior probability" OR "Posterior distribution" OR "Prior distribution" OR "Uninformative prior" OR prior OR "Credible interval" OR "Hierarchical model*" OR "Dynamic borrowing" OR "Predictive probability" ) ) ) AND ( ( TITLE-ABS-KEY ( ( "Clinical trial*" OR "drug evaluation*" OR "evaluation stud*" OR efficacy ) W/3 ( "phase 3" OR "phase III" OR "phase 4" OR "phase IV" ) ) ) OR ( TITLE-ABS-KEY ( ( platform* OR complet* OR adaptive* ) AND trial* ) ) OR ( ( ( "phase 3" OR "phase III" OR "phase 4" OR "phase IV" ) W/3 ( study OR studies OR trial* ) ) ) ) ) AND NOT ( TITLE-ABS-KEY ( ( "Clinical trial*" OR "drug evaluation*" OR "evaluation stud*" OR efficacy ) W/3 ( "phase 1" OR "phase I" ) ) ) AND PUBYEAR > 2009 AND PUBYEAR < 2025 AND ( LIMIT-TO ( DOCTYPE , "ar" ) ) AND ( LIMIT-TO ( LANGUAGE , "English" ) ) AND ( LIMIT-TO ( EXACTKEYWORD , "Humans" ) ) AND ( LIMIT-TO ( SUBJAREA , "MEDI" ) )

## Supplementary Table 3. Full List of studies included in this review

| # | Title | Author | Outcome | SSD Method |
|---|---|---|---|---|
| (1) | Adjunctive Intravenous Argatroban or Eptifibatide for Ischemic Stroke | Adeoye | continuous | Hybrid |
| (2) | Neoadjuvant Trebananib plus Paclitaxel-based Chemotherapy for Stage II/III Breast Cancer in the Adaptively Randomized I-SPY2 Trial-Efficacy and Biomarker Discovery. | Albain | binary | No justification |
| (3) | A randomized phase III study of pretransplant conditioning for AML/MDS with fludarabine and once daily IV busulfan +/- clofarabine in allogeneic stem cell transplantation | Andersson | survival | Frequentist |
| (4) | Effect of Hydrocortisone on Mortality and Organ Support in Patients With Severe COVID-19: The REMAP-CAP COVID-19 Corticosteroid Domain Randomized Clinical Trial | Angus | ordinal | Hybrid |
| (5) | Lopinavir-ritonavir and hydroxychloroquine for critically ill patients with COVID-19: REMAP-CAP randomized controlled trial | Arabi | ordinal | Hybrid |
| (6) | Therapeutic Anticoagulation with Heparin in Noncritically Ill Patients with Covid-19 | ATTACC investigators (Lawler) | ordinal | Hybrid |
| (7) | Safety and efficacy of neublastin in painful lumbosacral radiculopathy: a randomized, double-blinded, placebo-controlled phase 2 trial using Bayesian adaptive design (the SPRINT trial) | Backonja | continuous | Hybrid |
| (8) | Patient Assisted Intervention for Neuropathy: Comparison of Treatment in Real Life Situations (PAIN-CONTRoLS): Bayesian Adaptive Comparative Effectiveness Randomized Trial | Barohn | ordinal | Hybrid |
| (9) | Effect of P2Y12 Inhibitors on Organ Support-Free Survival in Critically Ill Patients Hospitalized for COVID-19: A Randomized Clinical Trial. | Berger | ordinal | Hybrid |
| (10) | Effect of P2Y12 Inhibitors on Survival Free of Organ Support Among Non-Critically Ill Hospitalized Patients With COVID-19: A Randomized Clinical Trial | Berger | ordinal | Hybrid |
| (11) | Demonstration of a 'leapfrog' randomized controlled trial as a method to accelerate the development and optimization of psychological interventions | Blackwell | continuous | Hybrid |
| (12) | The effects on self-efficacy, motivation and perceived barriers of an intervention targeting physical activity and sedentary behaviours in office workers: a cluster randomized control trial | Blom | continuous | Frequentist |
| (13) | Effect of Antiplatelet Therapy on Survival and Organ Support-Free Days in Critically Ill Patients With COVID-19: A Randomized Clinical Trial | Bradbury | ordinal | Hybrid |
| (14) | A Phase II randomized study of galunisertib monotherapy or galunisertib plus lomustine compared with lomustine monotherapy in patients with recurrent glioblastoma | Brandes | survival | Hybrid |
| (15) | Comparison of two chemotherapy regimens in patients with newly diagnosed Ewing sarcoma (EE2012): an open-label, randomised, phase 3 trial | Brennan | survival | Hybrid |
| (16) | Mycophenolate Mofetil Versus Cyclophosphamide for Remission Induction in Childhood Polyarteritis Nodosa: An Open-Label, Randomized, Bayesian Noninferiority Trial | Brogan | binary | Bayesian |
| (17) | First trimester anatomy ultrasound for patients with obesity: a randomized controlled trial. | Buskmiller | binary | No justification |
| (18) | Molnupiravir plus usual care versus usual care alone as early treatment for adults with COVID-19 at increased risk of adverse outcomes (PANORAMIC): an open-label, platform-adaptive randomised controlled trial | Butler | binary | Hybrid |

| (19) | Oseltamivir plus usual care versus usual care for influenza-like illness in primary care: an open-label, pragmatic, randomised controlled trial | Butler | survival | Hybrid |
|---|---|---|---|---|
| (20) | An Inhaled PI3Kdelta Inhibitor Improves Recovery in Acutely Exacerbating COPD Patients: A Randomized Trial | Cahn | continuous | Frequentist |
| (21) | Early and late preterm birth rates in participants adherent to randomly assigned high dose docosahexaenoic acid (DHA) supplementation in pregnancy | Carlson | binary | Hybrid |
| (22) | Efficacy of levetiracetam, fosphenytoin, and valproate for established status epilepticus by age group (ESETT): a double-blind, responsive-adaptive, randomised controlled trial | Chamberlain | binary | Hybrid |
| (23) | MK-2206 and Standard Neoadjuvant Chemotherapy Improves Response in Patients With Human Epidermal Growth Factor Receptor 2-Positive and/or Hormone Receptor-Negative Breast Cancers in the I-SPY 2 Trial | Chien | binary | Hybrid |
| (24) | Effect of anakinra versus usual care in adults in hospital with COVID-19 and mild-to-moderate pneumonia (CORIMUNO-ANA-1): a randomised controlled trial | CORIMUNO-19 Collaborative group (Tharaux) | joint | Hybrid |
| (25) | Sarilumab in adults hospitalised with moderate-to-severe COVID-19 pneumonia (CORIMUNO-SARI-1): An open-label randomised controlled trial | CORIMUNO-19 Collaborative group (Mariette) | joint | Hybrid |
| (26) | Oral steroids for reducing kidney scarring in young children with febrile urinary tract infections: the contribution of Bayesian analysis to a randomized trial not reaching its intended sample size | Da Dalt | binary | Bayesian |
| (27) | Electric Field Navigated 1-Hz rTMS for Poststroke Motor Recovery: The E-FIT Randomized Controlled Trial. | Edwards | binary | Frequentist |
| (28) | Effect of Convalescent Plasma on Organ Support-Free Days in Critically Ill Patients With COVID-19: A Randomized Clinical Trial | Estcourt | ordinal | Hybrid |
| (29) | Report of the first seven agents in the I-SPY COVID trial: a phase 2, open label, adaptive platform randomised controlled trial | Files | joint | Hybrid |
| (30) | Namilumab or infliximab compared with standard of care in hospitalised patients with COVID-19 (CATALYST): a randomised, multicentre, multi-arm, multistage, open-label, adaptive, phase 2, proof-of-concept trial | Fisher | continuous | Hybrid |
| (31) | Coronary sinus reducer for the treatment of refractory angina (ORBITA-COSMIC): a randomised, placebo-controlled trial. | Foley | joint | Frequentist |
| (32) | Efficacy and Safety of Toreforant, a Selective Histamine H4 Receptor Antagonist, for the Treatment of Moderate-to-Severe Plaque Psoriasis: Results from a Phase 2 Multicenter, Randomized, Double-blind, Placebo-controlled Trial | Frankel | binary | No justification |
| (33) | Effects of suboptimal adherence of CPAP therapy on symptoms of obstructive sleep apnoea: a randomised, double-blind, controlled trial | Gaisl | continuous | Frequentist |
| (34) | Outpatient convalescent plasma therapy for high-risk patients with early COVID-19: a randomized placebo-controlled trial | Gharbharan | ordinal | Frequentist |
| (35) | Efficacy and safety of bimekizumab as add-on therapy for rheumatoid arthritis in patients with inadequate response to certolizumab pegol: a proof-of-concept study | Glatt | continuous | Hybrid |
| (36) | Therapeutic Anticoagulation with Heparin in Critically Ill Patients with Covid-19 | Goligher | ordinal | Hybrid |
| (37) | Interleukin-6 Receptor Antagonists in Critically Ill Patients with Covid-19 | Gordon | ordinal | Hybrid |

| (38) | A Randomized, Open-Label Phase 2 Study of the CXCR4 Inhibitor LY2510924 in Combination with Sunitinib Versus Sunitinib Alone in Patients with Metastatic Renal Cell Carcinoma (RCC) | Hainsworth | survival | Hybrid |
|---|---|---|---|---|
| (39) | Rilonacept for colchicine-resistant or -intolerant familial Mediterranean fever: a randomized trial | Hashkes | count | Frequentist |
| (40) | Effect of Interleukin-6 Receptor Antagonists in Critically Ill Adult Patients with COVID-19 Pneumonia: two Randomized Controlled Trials of the CORIMUNO-19 Collaborative Group | Hermine | joint | Hybrid |
| (41) | Simvastatin in Critically Ill Patients with Covid-19. | Hills | ordinal | Hybrid |
| (42) | Predictors of Response to Web-Based Cognitive Behavioral Therapy With High-Intensity Face-to-Face Therapist Guidance for Depression: A Bayesian Analysis | Hoifodt | continuous | No justification |
| (43) | Inhibition of the KCa2 potassium channel in atrial fibrillation: a randomized phase 2 trial | Holst | binary | Frequentist |
| (44) | Effectiveness of a Digital Inhaler System for Patients With Asthma: A 12-Week, Open-Label, Randomized Study (CONNECT1) | Hoyte | binary | Frequentist |
| (45) | Effectiveness of casirivimab and imdevimab, and sotrovimab during Delta variant surge: a prospective cohort study and comparative effectiveness randomized trial | Huang | ordinal | No justification |
| (46) | Biodegradable polymer sirolimus-eluting stents versus durable polymer everolimus-eluting stents in patients with ST-segment elevation myocardial infarction (BIOSTEMI): a single-blind, prospective, randomised superiority trial | Iglesias | binary | Hybrid |
| (47) | Angiotensin receptor blockers for the treatment of covid-19: pragmatic, adaptive, multicentre, phase 3, randomised controlled trial | Jardine | ordinal | Hybrid |
| (48) | CGRP blockade by galcanezumab was not associated with reductions in signs and symptoms of knee osteoarthritis in a randomized clinical trial | Jin | continuous | Hybrid |
| (49) | Effect of Levocarnitine vs Placebo as an Adjunctive Treatment for Septic Shock: The Rapid Administration of Carnitine in Sepsis (RACE) Randomized Clinical Trial | Jones | joint | Hybrid |
| (50) | Personalized Research on Diet in Ulcerative Colitis and Crohn's Disease: A Series of N-of-1 Diet Trials | Kaplan | not specified | Frequentist |
| (51) | Randomized Trial of Three Anticonvulsant Medications for Status Epilepticus | Kapur | binary | Hybrid |
| (52) | Efficacy and Safety of RBX2660 in PUNCH CD3, a Phase III, Randomized, Double-Blind, Placebo-Controlled Trial with a Bayesian Primary Analysis for the Prevention of Recurrent Clostridioides difficile Infection | Khanna | binary | Frequentist |
| (53) | A Randomised -Controlled Phase 2 trial of Molnupiravir in Unvaccinated and Vaccinated Individuals with Early SARS-CoV-2 | Khoo | survival | Hybrid |
| (54) | Unreamed Intramedullary Nailing Versus External Fixation for the Treatment of Open Tibial Shaft Fractures in Uganda: A Randomized Clinical Trial | Kisitu | continuous | No justification |
| (55) | Early convalescent plasma for high-risk outpatients with covid-19 | Korley | binary | Hybrid |
| (56) | Effect of sodium-glucose co-transporter-2 inhibitors on survival free of organ support in patients hospitalised for COVID-19 (ACTIV-4a): a pragmatic, multicentre, open-label, randomised, controlled, platform trial. | Kosiborod | ordinal | No justification |
| (57) | Comparison of vestipitant with ondansetron for the treatment of breakthrough postoperative nausea and vomiting after failed prophylaxis with ondansetron | Kranke | binary | No justification |
| (58) | A Randomised, Double Blind, Placebo-Controlled Pilot Study of Oral Artesunate Therapy for Colorectal Cancer | Krishna | binary | Frequentist |
| (59) | Use of covid-19 convalescent plasma to treat patients admitted to hospital for covid-19 with or without underlying immunodeficiency: open label, randomised clinical trial | Lacombe | binary | Hybrid |
| (60) | Pulmonary Vein Isolation With or Without Left Atrial Appendage Ligation in Atrial Fibrillation: The aMAZE Randomized Clinical Trial. | Lakkireddy | binary | Hybrid |
| (61) | COVID-19 Immunologic Antiviral Therapy With Omalizumab (CIAO)-a Randomized Controlled Clinical Trial | Le | binary | Hybrid |
| (62) | Hybrid-control arm construction using historical trial data for an early-phase, randomized controlled trial in metastatic colorectal cancer | Li | binary | No justification |
| (63) | Bayesian Adaptive Randomization Trial of Passive Scattering Proton Therapy and Intensity-Modulated Photon Radiotherapy for Locally Advanced Non-Small-Cell Lung Cancer | Liao | survival | Hybrid |
| (64) | A Randomized Double-Blinded Placebo Controlled Trial of Clazakizumab for the Treatment of COVID-19 Pneumonia With Hyperinflammation | Lonze | ordinal | Hybrid |
| (65) | High-Flow Nasal Oxygen vs Noninvasive Ventilation in Patients With Acute Respiratory Failure: The RENOVATE Randomized Clinical Trial | Maia | binary | Hybrid |
| (66) | The comparative effectiveness of COVID-19 monoclonal antibodies: A learning health system randomized clinical trial | McCreary | ordinal | No justification |
| (67) | Anticoagulation Strategies in Non-Critically Ill Patients with Covid-19. | McQuilten | binary | Hybrid |
| (68) | Unconditional cash transfers for clinical and economic outcomes among HIV-Affected Ugandan households | Mills | joint | Hybrid |
| (69) | Metronomic chemotherapy plus anti-PD-1 in metastatic breast cancer: a Bayesian adaptive randomized phase 2 trial. | Mo | binary | No justification |
| (70) | Infliximab versus interferon-alpha in the treatment of Behcet's Syndrome: Clinical data from the BIO-Behcet's randomised controlled trial | Moots | continuous | Bayesian |
| (71) | Effect of Pembrolizumab Plus Neoadjuvant Chemotherapy on Pathologic Complete Response in Women With Early-Stage Breast Cancer: An Analysis of the Ongoing Phase 2 Adaptively Randomized I-SPY2 Trial | Nanda | binary | Frequentist |
| (72) | The effectiveness of multi-component interventions targeting physical activity or sedentary behaviour amongst office workers: a three-arm cluster randomised controlled trial | Nooijen | continuous | Frequentist |
| (73) | Locus coeruleus integrity and the effect of atomoxetine on response inhibition in Parkinson's disease | O'Callaghan | continuous | No justification |
| (74) | NUTRARET: effect of 2-Year Nutraceutical Supplementation on Redox Status and Visual Function of Patients With Retinitis Pigmentosa: a Randomized, Double-Blind, Placebo-Controlled Trial | Olivares-Gonzalez | continuous | No justification |
| (75) | Topical Ocular Anti-TNFÎ± Agent Licaminlimab in the Treatment of Acute Anterior Uveitis: A Randomized Phase II Pilot Study | Pasquali | binary | No justification |
| (76) | Safety and Efficacy of the BNT162b2 mRNA Covid-19 Vaccine | Polack | binary | Hybrid |
| (77) | Trial of Early Minimally Invasive Removal of Intracerebral Hemorrhage | Pradilla | continuous | Hybrid |
| (78) | TG4010 immunotherapy and first-line chemotherapy for advanced non-small-cell lung cancer (TIME): results from the phase 2b part of a randomised, double-blind, placebo-controlled, phase 2b/3 trial | Quoix | survival | Hybrid |
| (79) | Treating intrusive memories after trauma in healthcare workers: a Bayesian adaptive randomised trial developing an imagery-competing task intervention | Ramineni | count | Frequentist |
| (80) | Effect of Weekly Paclitaxel With or Without Bevacizumab on Progression-Free Rate Among Patients With Relapsed Ovarian Sex Cord-Stromal Tumors: The ALIENOR/ENGOT-ov7 Randomized Clinical Trial | Ray-Coquard | binary | Hybrid |
| (81) | Immunomodulatory therapy in children with paediatric inflammatory multisystem syndrome temporally associated with SARS-CoV-2 (PIMS-TS, MIS-C; RECOVERY): a randomised, controlled, open-label, platform trial | RECOVERY Collaborative Group (Faust) | count | No justification |
| (82) | Effect of spirulina on risk of hospitalization among patients with COVID-19: the TOGETHER randomized trial. | Reis | binary | Frequentist |
| (83) | Early Treatment with Pegylated Interferon Lambda for Covid-19 | Reis | binary | Frequentist |
| (84) | Oral Fluvoxamine With Inhaled Budesonide for Treatment of Early-Onset COVID-19 : A Randomized Platform Trial | Reis | binary | Hybrid |
| (85) | Effect of Early Treatment with Ivermectin among Patients with Covid-19 | Reis | binary | Hybrid |
| (86) | Effect of early treatment with fluvoxamine on risk of emergency care and hospitalisation among patients with COVID-19: the TOGETHER randomised, platform clinical trial | Reis | binary | Hybrid |
| (87) | Effect of early treatment with metformin on risk of emergency care and hospitalization among patients with COVID-19: the TOGETHER randomized platform clinical trial | Reis | binary | Hybrid |
| (88) | Effect of Angiotensin-Converting Enzyme Inhibitor and Angiotensin Receptor Blocker Initiation on Organ Support-Free Days in Patients Hospitalized With COVID-19: A Randomized Clinical Trial | REMAP-CAP committee (Lawler) | ordinal | Hybrid |
| (89) | Trial of a Preferential Phosphodiesterase 4B Inhibitor for Idiopathic Pulmonary Fibrosis | Richeldi | continuous | Hybrid |
| (90) | The Effects of Opt-out vs Opt-in Tobacco Treatment on Engagement, Cessation, and Costs: A Randomized Clinical Trial | Richter | binary | Hybrid |
| (91) | Adaptive Randomization of Veliparib-Carboplatin Treatment in Breast Cancer | Rugo | binary | Hybrid |
| (92) | A nebulised antitumour necrosis factor receptor-1 domain antibody in patients at risk of postoperative lung injury: A randomised, placebo-controlled pilot study | Ryan | continuous | Hybrid |
| (93) | Infliximab versus Cyclophosphamide for Severe Behcet's Syndrome. | Saadoun | binary | Frequentist |
| (94) | Effect of Different Doses of Galcanezumab vs Placebo for Episodic Migraine Prevention: A Randomized Clinical Trial | Skljarevski | continuous | Hybrid |
| (95) | Effect of the P-Selectin Inhibitor Crizanlizumab on Survival Free of Organ Support in Patients Hospitalized for COVID-19: A Randomized Controlled Trial. | Solomon | ordinal | No justification |
| (96) | Effect of Intravitreal Aflibercept vs Laser Photocoagulation on Treatment Success of Retinopathy of Prematurity: The FIREFLEYE Randomized Clinical Trial | Stahl | binary | Hybrid |
| (97) | Randomized Study Comparing a Reusable Morcellator with a Resectoscope in the Hysteroscopic Treatment of Uterine Polyps: The RESMO Study | Stoll | continuous | Frequentist |
| (98) | Effect of Mexiletine on Muscle Stiffness in Patients With Nondystrophic Myotonia Evaluated Using Aggregated N-of-1 Trials | Stunnenberg | continuous | Hybrid |
| (99) | Sorafenib in Molecularly Selected Cancer Patients: Final Analysis of the MOST-Plus Sorafenib Cohort | Tredan | binary | Bayesian |
| (100) | No add-on effect of tDCS on fatigue and depression in chronic stroke patients: A randomized sham-controlled trial combining tDCS with computerized cognitive training | Ulrichsen | joint | No justification |
| (101) | Safety and efficacy of secukinumab in patients with giant cell arteritis (TitAIN): a randomised, double-blind, placebo-controlled, phase 2 trial. | Venhoff | binary | Hybrid |
| (102) | Multicenter, randomized, open-label Phase II study comparing S-1 alternate-day oral therapy with the standard daily regimen as a first-line treatment in patients with unresectable advanced pancreatic cancer | Yamaue | survival | Hybrid |
| (103) | Advanced reperfusion strategies for patients with out-of-hospital cardiac arrest and refractory ventricular fibrillation (ARREST): a phase 2, single centre, open-label, randomised controlled trial | Yannopoulos | binary | Hybrid |
| (104) | Intramyocardial Injection of Mesenchymal Precursor Cells and Successful Temporary Weaning From Left Ventricular Assist Device Support in Patients With Advanced Heart Failure: A Randomized Clinical Trial | Yau | count | Hybrid |
| (105) | Thrombectomy for Stroke with Large Infarct on Noncontrast CT: The TESLA Randomized Clinical Trial | Yoo | continuous | Hybrid |
| (106) | Safety and efficacy of trehalose in amyotrophic lateral sclerosis (HEALEY ALS Platform Trial): an adaptive, phase 2/3, double-blind, randomised, placebo-controlled trialA1 | HEALEY | joint | Hybrid |
| (107) | Verdiperstat in Amyotrophic Lateral Sclerosis: Results From the Randomized HEALEY ALS Platform Trial | Andrews | joint | Hybrid |
| (108) | CNM-Au8 in Amyotrophic Lateral Sclerosis: The HEALEY ALS Platform Trial | Berry | joint | Hybrid |
| (109) | Pridopidine in Amyotrophic Lateral Sclerosis: The HEALEY ALS Platform Trial | Shefner | joint | Hybrid |
| (110) | Efficacy and Safety of Zilucoplan in Amyotrophic Lateral Sclerosis: A Randomized Clinical Trial | Paganoni | joint | Hybrid |
| (111) | Effect of hydrocortisone on mortality in patients with severe community-acquired pneumonia: The REMAP-CAP Corticosteroid Domain Randomized Clinical Trial | Annane | binary | Hybrid |
| (112) | Tocilizumab, sarilumab and anakinra in critically ill patients with COVID-19: a randomised, controlled, open-label, adaptive platform trial | Derde | ordinal | No Justification |
| (113) | Inhaled Fluticasone Furoate for Outpatient Treatment of Covid-19 | Boulware | survival | Hybrid |
| (114) | Metformin and Time to Sustained Recovery in Adults With COVID-19: The ACTIV-6 Randomized Clinical Trial | Bramante | survival | Hybrid |
| (115) | Fostamatinib for Hospitalized Adults With COVID-19 and Hypoxemia A Randomized Clinical Trial | Collins | count | Hybrid |
| (116) | Effect of Fluvoxamine vs Placebo on Time to Sustained Recovery in Outpatients With Mild to Moderate COVID-19: A Randomized Clinical Trial | McCarthy | survival | Hybrid |
| (117) | Effect of Ivermectin vs Placebo on Time to Sustained Recovery in Outpatients With Mild to Moderate COVID-19: A Randomized Clinical Trial | Naggie | survival | Hybrid |
| (118) | Effect of Higher-Dose Ivermectin for 6 Days vs Placebo on Time to Sustained Recovery in Outpatients With COVID-19: A Randomized Clinical Trial | Naggie | survival | Hybrid |
| (119) | Higher-Dose Fluvoxamine and Time to Sustained Recovery in Outpatients With COVID-19: The ACTIV-6 Randomized Clinical Trial | Stewart | survival | Hybrid |
| (120) | Azithromycin for community treatment of suspected COVID-19 in people at increased risk of an adverse clinical course in the UK (PRINCIPLE): a randomised, controlled, open-label, adaptive platform trial | Butler | multiple | Hybrid |
| (121) | Doxycycline for community treatment of suspected COVID-19 in people at high risk of adverse outcomes in the UK (PRINCIPLE): a randomised, controlled, open-label, adaptive platform trial | Butler | multiple | Hybrid |
| (122) | Colchicine for COVID-19 in the community (PRINCIPLE): a randomised, controlled, adaptive platform trial | Dorward | multiple | Hybrid |
| (123) | Ivermectin for COVID-19 in adults in the community (PRINCIPLE): An open, randomised, controlled, adaptive platform trial of short- and longer-term outcomes | Hayward | multiple | Hybrid |
| (124) | Favipiravir for COVID-19 in adults in the community in PRINCIPLE, an open-label, randomised, controlled, adaptive platform trial of short- and longer-term outcomes | Hobbs | multiple | Hybrid |
| (125) | Inhaled budesonide for COVID-19 in people at high risk of complications in the community in the UK (PRINCIPLE): a randomised, controlled, open-label, adaptive platform trial | Yu | multiple | Hybrid |
| (126) | Efficacy, immunogenicity, and safety of an investigational maternal respiratory syncytial virus prefusion F protein-based vaccine | Banooni | binary | No Justification |
| (127) | Paclitaxel With Inhibitor of Apoptosis Antagonist, LCL161, for Localized Triple-Negative Breast Cancer, Prospectively Stratified by Gene Signature in a Biomarker-Driven Neoadjuvant Trial | Bardia | binary | No Justification |
| (128) | Challenges in Recruiting University Students for Web-Based Indicated Prevention of Depression and Anxiety: Results From a Randomized Controlled Trial (ICare Prevent) | Bolinski | continuous | No Justification |
| (129) | Nicotine e-cigarettes for smoking cessation following discharge from smoke-free inpatient alcohol and other drug withdrawal services: a pragmatic two-arm, single-blinded, parallel-group, randomised controlled trial | Bonevski | joint | Frequentist |
| (130) | Efficacy and safety of eliapixant in diabetic neuropathic pain and prediction of placebo responders with an exploratory novel algorithm: results from the randomized controlled phase 2a PUCCINI study | Bouhassira | continuous | Hybrid |
| (131) | Postnatal Exercise Partners Study (PEEPS): a pilot randomized trial of a dyadic physical activity intervention for postpartum mothers and a significant other | Carr | continuous | Hybrid |
| (132) | Pharmacist-Led Education Intervention for Adults With Allergic Rhinitis: A Randomized Clinical Trial | Chew | multiple | Frequentist |
| (133) | Treatment options to support the elimination of hepatitis C: an open-label, factorial, randomised controlled non-inferiority trial | Cooke | binary | Frequentist |
| (134) | Vaporized Nicotine Products for Smoking Cessation Among People Experiencing Social Disadvantage: A Randomized Clinical Trial | Courtney | binary | Frequentist |
| (135) | RSV Prefusion F Protein-Based Maternal Vaccine - Preterm Birth and Other Outcomes | Dieussaert | binary | Hybrid |
| (136) | Efficacy and safety of artemether-lumefantrine and dihydroartemisinin-piperaquine for the treatment of uncomplicated Plasmodium falciparum malaria and prevalence of molecular markers associated with artemisinin and partner drug resistance in Uganda | Ebong | survival | Frequentist |
| (137) | Efficacy and Safety of Eliapixant in Overactive Bladder: The 12-Week, Randomised, Placebo-controlled Phase 2a OVADER Study | Ewerton | continuous | Hybrid |
| (138) | Discharge intervention to improve outcomes and web-based portal engagement after stroke and transient ischaemic attack: A randomised controlled trial | Fakes | joint | Frequentist |
| (139) | The effect of a combined indomethacin and levonorgestrel-releasing intrauterine system on short-term postplacement bleeding profile: a randomized proof-of-concept trial | Fels | count | Hybrid |
| (140) | Assessment of Efficacy and Safety of Zagotenemab: Results From PERISCOPE-ALZ, a Phase 2 Study in Early Symptomatic Alzheimer Disease | Fleisher | continuous | Hybrid |
| (141) | Cannabidiol for the treatment of cannabis use disorder: a phase 2a, double-blind, placebo-controlled, randomised, adaptive Bayesian trial | Freeman | joint | Frequentist |
| (142) | Efficacy and Safety of Bimekizumab in Moderate to Severe Hidradenitis Suppurativa: A Phase 2, Double-blind, Placebo-Controlled Randomized Clinical Trial | Glatt | binary | Hybrid |

| (143) | Establishing the lowest penicillin concentration to prevent pharyngitis due to Streptococcus pyogenes using a human challenge model (CHIPS): a randomised, double-blind, placebo-controlled trial | Hla | binary | Bayesian |
|---|---|---|---|---|
| (144) | Randomized controlled trial of early aerobic exercise following sport-related concussion: Progressive percentage of age-predicted maximal heart rate versus usual care | Hutchison | survival | Frequentist |
| (145) | Individualizing non-steroidal anti-inflammatory drug therapy in axial spondyloarthritis: N-of-1 trials with Bayesian analysis | Hwang | categorical and continuous | No Justification |
| (146) | Ivacaftor in People with Cystic Fibrosis and a 3849+10kb C->T or D1152H Residual Function Mutation | Kerem | continuous | Frequentist |
| (147) | Efficacy of the Nav1.7 blocker PF-05089771 in a randomised, placebo-controlled, double-blind clinical study in subjects with painful diabetic peripheral neuropathy | McDonnell | continuous | Hybrid |
| (148) | Erenumab for Chronic Cluster Headache: A Randomized Clinical Trial | Mecklenburg | continuous | Hybrid |
| (149) | Lemborexant, A Dual Orexin Receptor Antagonist (DORA) for the Treatment of Insomnia Disorder: Results From a Bayesian, Adaptive, Randomized, Double-Blind, Placebo-Controlled Study | Murphy | joint | Hybrid |
| (150) | Delivering therapy over telephone in a humanitarian setting: a pilot randomized controlled trial of common elements treatment approach (CETA) with Syrian refugee children in Lebanon | Pluess | continuous | Frequentist |
| (151) | Randomized 52-week phase 2 trial of albiglutide versus placebo in adult patients with newly diagnosed type 1 diabetes | Pozzilli | continuous | Hybrid |
| (152) | Strategies to Promote Resiliency: A Randomized Embedded Multifactorial Adaptative Platform (REMAP) Clinical Trial to Study Interventions to Improve Recovery After Surgery in High-Risk Patients. | Reitz | binary | Hybrid |
| (153) | Soccer and Vocational Training are Ineffective Delivery Strategies to Prevent HIV and Substance Abuse by Young, South African Men: A Cluster Randomized Controlled Trial | Rotheram-Borus | joint | Frequentist |
| (154) | On-Demand Sildenafil as a Treatment for Raynaud Phenomenon: A Series of n-of-1 Trials | Roustit | joint | No Justification |
| (155) | Contingency management plus acceptance and commitment therapy for initial cocaine abstinence: Results of a sequential multiple assignment randomized trial (SMART | Schmitz | binary | Hybrid |
| (156) | The effectiveness of medical nutrition therapy for people at moderate to high risk of cardiovascular disease in an Australian rural primary care setting: 12-month results from a pragmatic cluster randomised controlled trial | Schumacher | continuous | Frequentist |
| (157) | Effectiveness of a Mobile Phone-Delivered Multiple Health Behavior Change Intervention (LIFE4YOUth) in Adolescents: Randomized Controlled Trial | Seiterö | joint | Hybrid |
| (158) | Randomized trial of BCG in healthcare workers to reduce absenteeism during the COVID-19 pandemic in sub-Saharan Africa | Silva | count | Frequentist |
| (159) | Effects of Lumacaftor/Ivacaftor on Cystic Fibrosis Disease Progression in Children 2 through 5 Years of Age Homozygous for F508del-CFTR: A Phase 2 Placebo-controlled Clinical Trial | Stahl | binary | Hybrid |
| (160) | Facilitating treatment initiation and reproductive care postpartum to prevent substance-exposed pregnancies: A randomized bayesian pilot trial | Stotts | joint | Hybrid |
| (161) | Citalopram for treatment of cocaine use disorder: A Bayesian drop-the-loser randomized clinical trial | Suchting | count | Bayesian |
| (162) | A randomized, double-blind, phase 2b proof-of-concept clinical trial in early Alzheimer's disease with lecanemab, an anti-Abeta protofibril antibody | Swanson | continuous | Bayesian |

| (163) | Optimal strategies to improve uptake of and adherence to HIV prevention among young people at risk for HIV acquisition in the USA (ATN 149): a randomised, controlled, factorial trial | Swendeman | multiple | Frequentist |
|---|---|---|---|---|
| (164) | Low dose dexamethasone as treatment for women with heavy menstrual bleeding: A response-adaptive randomised placebo-controlled dose-finding parallel group trial (DexFEM) | Warner | continuous | Frequentist |